# Machine learning correction of satellite precipitation is governed by mechanism purity, not algorithmic complexity: a proof-of-concept study in Hunan, China, with pre-registered cross-regional validation

Yi Xu*

*School of Information Engineering, Hunan Industry Polytechnic, Changsha, Hunan, China*

*ORCID: 0009-0002-3789-8136*

**Corresponding author: Yi.Xu.Prof@outlook.com*

## Abstract

Satellite precipitation products such as IMERG exhibit biases that vary with terrain, season, and precipitation regime, leaving the applicability boundaries of machine learning correction unclear. This study proposes the Terrain–Moisture–Intensity (TMI) framework, centered on mechanism purity, extending the correction problem from purely algorithmic optimization to physical consistency diagnosis. A proof-of-concept study in Hunan Province employs IMERG V07, SRTM DEM, and ERA5 variables (tcwv, u10, v10). Ablation results indicate that, under the conditions of this study, terrain–moisture relationships are predominantly additive: RF-Full yields merely +0.001 $R^2$ gain over LR-Full, while bias rises to 1.282 mm $d^{-1}$; MAE decreases by approximately 14%, reflecting a trade-off between tail-fitting improvement and mean shift. SHAP diagnostics identify three categories of boundaries. Spatially, Central Hunan exhibits significant degradation ($R^2$ = 0.133) despite strong variable activation, consistent with mechanism fragmentation induced by mixed terrain. Temporally, u10 undergoes directional reversal between summer and spring (+0.096 to −0.156), presenting "silent failure." Extreme precipitation (≥50 mm $d^{-1}$) approximates a mechanism saturation frontier rather than isolated out-of-distribution samples, with DEM showing the largest relative amplification in SHAP disorder (+150%). The results demonstrate that machine learning correction performance is primarily constrained by mechanism purity. A pre-registered cross-regional test (Hunan, Guangxi, Guangdong) confirms this screening capability out of sample: a priori coherence proxies predict correction efficiency with a mean absolute error of 2.6 percentage points, while the transfer-versus-retraining contrast separates mechanism mismatch (coastal Guangdong) from portability (Guangxi), establishing the framework as a validated applicability screen.

**Keywords:** *Satellite precipitation correction; Random forest; SHAP; Mechanism purity; TMI framework; IMERG*

## 1. Introduction

Accurate precipitation estimation remains one of the most persistent challenges in atmospheric remote sensing. Ground-based gauge networks, while serving as the observational benchmark, suffer from spatial sparsity and uneven distribution, particularly over complex terrain and oceanic regions (Kidd et al., 2017). The Global Precipitation Measurement (GPM) mission, launched in 2014, represents the current state of the art in satellite precipitation remote sensing (Hou et al., 2014; Skofronick-Jackson et al., 2017), with its Integrated Multi-satellitE Retrievals for GPM (IMERG) product providing quasi-global coverage at 0.1° spatial and 30-minute temporal resolution. Despite substantial algorithmic advances, IMERG—and satellite precipitation products generally—exhibit systematic biases that vary with topography, season, and precipitation regime, limiting their utility for atmospheric process studies, climate monitoring, and numerical weather prediction (Tang et al., 2016; Tapiador et al., 2012; Derin and Yilmaz, 2014; Tang et al., 2020; Lei et al., 2022; Zhou et al., 2020).

The sources of IMERG error are multifaceted and physically grounded. Microwave-based retrieval struggles with orographic enhancement and shallow warm-rain processes, while infrared estimation introduces cold-cloud-top temperature biases that misrepresent convective intensity. Over complex terrain such as the subtropical monsoon regions of southern China, these errors compound: topographic modulation of precipitation interacts with moisture advection and convective organization in ways that retrieval algorithms, designed for large-scale stratiform systems, fail to capture. The result is a structured, spatially heterogeneous bias field that demands correction beyond simple mean adjustment (Derin and Yilmaz, 2014; Yan et al., 2022; Lei et al., 2022; Chen et al., 2020).

Traditional bias correction methods—linear regression, quantile mapping, and distribution-based approaches—are computationally efficient and interpretable but assume stationary, often linear, error structures (Chen et al., 2021). In regions where terrain heterogeneity fragments precipitation mechanisms and seasonal circulation shifts reconfigure moisture pathways, these assumptions break down. Machine learning algorithms, by contrast, offer the theoretical capacity to learn nonlinear, multivariate relationships between satellite estimates and ground truth without prespecified functional forms (Breiman, 2001; Liaw and Wiener, 2002; Pedregosa et al., 2011).

The random forest (RF) algorithm has emerged as a particularly popular choice for precipitation correction. As an ensemble of decision trees trained on bootstrap-aggregated samples, RF handles high-dimensional predictor spaces, captures interaction effects, and provides built-in measures of variable importance (Lin et al., 2022; Liu et al., 2022). Applications of machine learning to precipitation correction have reported substantial improvements in correlation and error metrics across diverse climatic regimes, using random forest (Liu et al., 2022; Bisht et al., 2025) as well as other nonlinear architectures such as gradient boosting and deep neural networks (Li et al., 2023; Dao et al., 2025). Systematic large-scale comparisons have likewise ranked candidate correction algorithms by predictive accuracy (Papacharalampous et al., 2023a, 2023b). Yet these studies typically evaluate performance through aggregate statistics—$R^2$, RMSE, MAE—without interrogating the physical mechanisms that govern when, where, and why correction succeeds or fails. This opacity matters: a model that improves $R^2$ by redistributing error from pointwise residuals toward the distribution mean may degrade physical utility despite superior summary metrics—a trade-off quantified directly in Section 3.1.

We advance a framework that reframes ML correction not as an algorithmic optimization problem but as a physical mechanism alignment problem. The core hypothesis is that ML-based correction is valid only to the extent that the training domain exhibits mechanism purity—the dominance of a single, coherent precipitation process within a given spatiotemporal window. Where purity is high (orographic convection over coherent terrain, stable seasonal circulation, moderate precipitation intensities), the learned relationships transfer robustly. Where purity degrades (mixed topography, transitional seasons, extreme intensities), variables shift from cooperative predictors to conflicting signals, and correction collapses.

To operationalize this hypothesis, we introduce the Terrain–Moisture–Intensity (TMI) framework, a three-dimensional diagnostic system that decomposes correction failure into spatial, temporal, and intensity boundaries. Each boundary is quantified through SHAP (SHapley Additive exPlanations; Lundberg and Lee, 2017) analysis, which opens the RF black box to reveal not merely which variables matter, but how their directional contributions and interaction dynamics vary across the TMI space. The TMI framework thus transforms ML correction from a predictive endpoint into a diagnostic instrument for probing the physical limits of satellite retrieval.

As a proof of concept, this study applies the TMI framework to summer precipitation correction over Hunan Province, a subtropical monsoon region characterized by complex topography (Nanling and Xuefeng mountain ranges, Dongting Lake plain), sharp seasonal transitions (Mei-yu frontal to orographic convective regimes), and frequent extreme precipitation events. Our objectives are threefold: (i) quantify overall correction performance through systematic ablation experiments that decompose the contribution of IMERG base signal, terrain elevation, and atmospheric reanalysis variables to correction gain; (ii) diagnose mechanism boundaries using SHAP analysis to identify where (spatially), when (temporally), and at what intensities the learned correction relationships degrade; and (iii) formalize the mechanism purity principle by demonstrating that correction collapse across all three TMI dimensions reflects a single underlying phenomenon: the fragmentation of a coherent precipitation mechanism into competing physical processes. To support replication and extension of the framework, the complete training, ablation, and cross-regional validation pipeline—including the shared coherence-proxy implementation—is openly released on GitHub and archived on Zenodo (see Code availability).

## 2. Data and methods

### *2.1. Study area*

Hunan Province (24°38′–30°08′N, 108°47′–114°15′E; ~211,800 km$^2$) spans south-central China and provides a suitable testbed for mechanism purity analysis given its topographic diversity and climatological complexity (Chen et al., 2018). The province is bounded by the Wuling and Xuefeng Mountains to the west and northwest, the Nanling Range to the south, and opens northward into the Dongting Lake plain and the middle Yangtze River basin. Elevation ranges from below 50 m in the northern lake district to over 2,000 m in the western mountain ranges, creating four distinct topographic regimes that govern precipitation mechanisms.

The climate is dominated by the East Asian subtropical monsoon. Summer (June–August) is characterized by two distinct phases: the Mei-yu (plum rain) period in June to early July, when quasi-stationary frontal systems produce widespread stratiform precipitation, and the post-Mei-yu period from mid-July to August, when orographic convection and typhoon remnants dominate. This seasonal dichotomy provides a natural experiment for testing temporal mechanism portability.

We divide Hunan Province into four subregions for spatial analysis: West Hunan (Wuling–Xuefeng foothills, topographic convection), Central Hunan (hill–plain mosaic, mixed mechanisms), North Hunan (Dongting Lake Plain, stratiform-dominated), and South Hunan (Nanling Range, orographic enhancement). This partition is based on topographic coherence and precipitation climatology rather than administrative boundaries, ensuring that each subregion approximates a single dominant mechanism. A K-means++ clustering evaluation (K = 2–8; Arthur and Vassilvitskii, 2007) further supports the optimality of the four-partition scheme from a data-driven perspective: the composite ranking of the silhouette coefficient, Calinski–Harabasz index, and Davies–Bouldin index indicates that K = 4 achieves the best balance (see Supplementary Table S1).

### *2.2. Data*

Observed precipitation uses the CN05.1 gridded daily precipitation dataset (Wu and Gao, 2013; Xu et al., 2009; Wu et al., 2017) as ground truth. Satellite precipitation estimates are from the GPM IMERG Version 07 product (Huffman et al., 2023). Auxiliary meteorological variables are from the ERA5 reanalysis (Hersbach et al., 2020; Copernicus Climate Change Service, 2023): total column water vapor (tcwv, kg $m^{-2}$), 10-m zonal wind (u10, m $s^{-1}$), and 10-m meridional wind (v10, m $s^{-1}$), selected to represent atmospheric moisture conditions and low-level moisture advection. Terrain is represented by the SRTM V3 Digital Elevation Model (DEM, native resolution 30 m; Big Earth Data Science Data Center, Chinese Academy of Sciences (CASEarth), 2025).

All variables are remapped to a common 0.25° grid (108.65°–114.35°E, 24.50°–30.30°N), whose extent is defined by the coverage of the pre-aggregated 0.25° DEM field and therefore differs slightly from the administrative boundary of Hunan Province. The 30 m SRTM DEM is first area-averaged to 0.25° during preprocessing; ERA5 continuous variables are then linearly interpolated to the target grid, while IMERG, observed precipitation, and the pre-aggregated DEM are aligned to it by nearest-neighbor assignment. A multivariate finite-value mask is applied, retaining zero-precipitation samples; these are kept to preserve the complete sample distribution and to avoid subjectivity in rain/no-rain threshold selection, so all skill metrics are evaluated over the full set of grid-days. The study domain covers Hunan Province grid cells, yielding 321 valid grid points.

Summer (June–August) data span 2016–2022, split by year into training (2016–2020) and testing (2021–2022) periods, comprising 147,660 and 59,064 grid-day samples, respectively. For seasonal transferability analysis, March–May 2016–2022 samples are additionally used under the same variable framework for comparative evaluation. The spring transferability analysis follows the same year-based split, with 2016–2020 used for training and 2021–2022 for evaluation.

For cross-regional validation (Sections 2.3.5 and 3.6), identical datasets are assembled for Guangxi and Guangdong provinces—CN05.1 observed precipitation, IMERG V07, ERA5 tcwv/u10/v10, and SRTM DEM—preprocessed with the same remapping, interpolation, and finite-value masking procedure on provincial 0.25° grids clipped to the administrative boundaries, yielding 329 (Guangxi) and 243 (Guangdong) valid grid cells over the same 2016–2022 summer period.

### *2.3. Methods*

#### 2.3.1. Random forest model and ablation design

We build a random forest regression model using scikit-learn (Pedregosa et al., 2011) with the following parameter configuration: n_estimators = 500, max_features = 1.0, remaining parameters at default (including bootstrap = True), and a fixed random seed (random_state = 42) for reproducibility; training uses parallel computation (n_jobs = -1). Out-of-bag (OOB) scoring is not enabled. We avoid hyperparameter optimization to prevent tuning artifacts and ensure comparability across model configurations.

RF-Full uses five predictors: IMERG precipitation, u10, v10, tcwv, and DEM. Samples are split by year: 2016–2020 for training and 2021–2022 for testing. The summer model uses June–August samples; the spring model (RF-Spring) uses March–May samples for seasonal transferability analysis under the unified variable set.

To decompose the incremental contribution of each predictor group, we design a six-step ablation sequence:

(1) Raw IMERG as baseline;
(2) IMERG-only RF (IMERG precipitation as the sole input);
(3) IMERG + DEM;
(4) IMERG + ERA5 (u10, v10, tcwv, without DEM);
(5) LR-Full (five-variable linear regression, as the linear reference);
(6) RF-Full (five-variable random forest). The comparison between steps (5) and (6) tests for mechanistic nonlinearity: if RF-Full substantially exceeds LR-Full, the terrain–moisture relationship contains exploitable nonlinear structure.

Additionally, we evaluate 2021 and 2022 summers independently to compare RF–LR algorithmic ranking and bias trajectories across years; and apply a five-day moving average to the daily mean bias series to suppress synoptic-scale noise and highlight intraseasonal regime transitions.

Model performance is quantified using the coefficient of determination ($R^2 = 1 - SSE/SST$, where SSE is the sum of squared errors and SST is the total sum of squares about the observed mean; $R^2$ is equivalent to the Nash–Sutcliffe efficiency and can take negative values, which indicate predictions worse than the observed climatological mean), root-mean-square error (RMSE), mean absolute error (MAE), and mean bias (Bias = estimate − observation, positive values indicating overestimation). All metrics are computed over grid-day samples of the test period.

### 2.3.2. SHAP interpretability analysis

We employ SHAP to explain model prediction contributions. For tree-based models, SHAP values are computed using TreeExplainer with the tree_path_dependent feature perturbation setting and approximate computation enabled (approximate = True) for efficiency.

From SHAP values, we derive three diagnostic metrics:

(1) Variable importance: quantified as the mean absolute SHAP value per feature for global ranking;
(2) Directional consistency: within each precipitation intensity stratum, measured by the Spearman correlation between feature values and SHAP values to characterize upward/downward correction direction; cross-season stability is assessed using the mean SHAP sign; SHAP dependence plots are used to assist in identifying nonlinear coupling between variables;
(3) SHAP disorder index: defined as the standard deviation ($\sigma$) of SHAP values within each intensity class, characterizing mechanistic dispersion; the escalation of $\sigma$ at extreme intensities is interpreted as a candidate signal of mechanistic saturation.

### 2.3.3. Stratification and statistical testing

We stratify all analyses by precipitation intensity class: light rain (0.1–10 mm $d^{-1}$), moderate rain (10–25 mm $d^{-1}$), heavy rain (25–50 mm $d^{-1}$), and torrential rain (≥50 mm $d^{-1}$). For subregional performance differences, bootstrap is applied to estimate 95% confidence intervals of $R^2$ differences (N = 1,000 resamples), with $\alpha = 0.05$ as the significance threshold. The current implementation targets the $R^2$ difference between Central Hunan and South Hunan. Because grid-day samples are spatially and temporally autocorrelated, these resampling-based intervals should be regarded as optimistic (lower-bound) estimates of the true uncertainty; block-bootstrap refinements are discussed in Section 4.4.

To test the domain-shift hypothesis for extreme precipitation, we apply t-SNE (van der Maaten and Hinton, 2008) in the five-dimensional feature space. The sampling strategy retains 300 non-torrential cases and includes all torrential cases (522 in this study, total n = 822). Based on the local neighborhood structure in the two-dimensional embedding, we determine whether torrential samples form relatively isolated clusters (supporting out-of-distribution failure) or extend continuously from the non-torrential domain (supporting saturation-frontier failure).

### 2.3.4. Operationalizing mechanism purity

We define mechanism purity formally. Let W denote a spatiotemporal analysis window (a region–season–intensity stratum) containing samples generated by one or more precipitation mechanisms, each inducing its own predictor–predictand mapping $f_k: X \rightarrow y$. Mechanism purity P(W) is the degree to which a single such mapping dominates W; it is maximal when all samples share one mechanism and degrades as samples mix incompatible mappings. Three distinctions delimit the concept. First, purity differs from signal-to-noise ratio: a window may contain a strong but heterogeneous signal (several mechanisms, each strong) or a weak but pure one; purity concerns the singularity of the mapping, not its amplitude. Second, purity differs from causal strength: a predictor may act strongly within each of two coexisting mechanisms whose directions conflict, so causal influence per se does not guarantee learnability. Third, purity differs from physical consistency: purity is a property of the data-generating window, whereas consistency is a property of model behavior within that window.

We operationalize purity at two stages. A priori, before any model fitting, regime coherence is quantified from observational climatology: here we use the mean daily spatial coherence of observed precipitation within each subregion, measured by Moran's I (Moran, 1950; Anselin, 1995), computed with queen contiguity on the 0.25° grid and averaged over test-period days, as a proxy for the spatial organization of precipitation mechanisms—organized regimes (coherent orographic or frontal systems) yield high values, whereas fragmented, mixed regimes yield low values. A posteriori, after model fitting, purity is diagnosed from learned behavior through SHAP directional consistency and the SHAP disorder index (Section 2.3.2). If purity genuinely constrains learnability, the a priori proxy—computed without any model output—must rank regions in the order of their subsequent correction skill; Section 3.3.4 tests this prediction.

### 2.3.5. Cross-regional validation design and pre-registration

Cross-regional validation covers Guangxi and Guangdong under the identical pipeline (Section 2.2) and the same 2016–2020 training / 2021–2022 testing split. Guangxi is stratified into a karst northwest subregion (longitude < 108°E or elevation > 500 m) and southeastern hills, isolating the fragmented karst plateau from the lower-relief southeast; Guangdong into coastal (< 22.7°N), inland-hill (22.7–24°N), and northern-mountain (≥ 24°N) subregions, separating the typhoon-influenced coastal plain, the inland hills, and the orographic Nanling belt. The design crosses two arms—direct transfer of a source-province model to each target province, testing mechanism portability, and local retraining on the target province, testing the learnability gate—within a leave-one-region-out scheme that rotates source and target roles across the three provinces, yielding the full 3 × 3 transfer matrix for both RF and LR. Correction efficiency (the percentage RMSE reduction relative to raw IMERG; Section 3.3.4) is computed on the common test period, with 95% confidence intervals from a day-block bootstrap (1,000 resamples) that preserves serial correlation. The a priori test is pre-registered: the linear coherence–efficiency rule is fitted on the four Hunan subregions only, applied to the Guangxi and Guangdong coherence proxies, and the five subregion predictions are written to a UTC-timestamped record before any Guangxi or Guangdong model is evaluated (Supplementary Table S8); predicted and realized efficiencies are then compared both continuously (absolute error) and categorically (low/medium/high classes).

# 3. Results

## *3.1. Main model performance: RF correction effect and variable contribution*

### 3.1.1. Visual diagnosis of scatter patterns

Raw IMERG exhibits severe systematic bias relative to observations ($R^2$ = −0.215, RMSE = 10.902 mm $d^{-1}$, MAE = 4.816 mm $d^{-1}$, Bias = 0.478 mm $d^{-1}$). As shown in Fig. 1a, IMERG scatters densely in the low-value region (<20 mm $d^{-1}$), while estimates for high observed values are substantially underestimated. The overall scatter cloud deviates markedly from the 1:1 reference line, indicating strong structural underestimation of intense precipitation in the raw product; this high-end underestimation coexists with a small positive domain-mean bias (+0.478 mm $d^{-1}$), reflecting compensation between high-intensity underestimation and low-intensity overestimation.

After RF-Full correction, the scatter cloud converges toward the 1:1 reference line overall (Fig. 1b), with $R^2$ improved from −0.215 to 0.344, RMSE reduced to 8.009 mm $d^{-1}$ (a decrease of ~26.5% relative to raw IMERG), and MAE decreased from 4.816 to 4.166 mm $d^{-1}$. The model achieves more stable pointwise error convergence in the light-to-moderate precipitation range, demonstrating that the TMI feature space effectively constrains correction errors in this regime.

However, significant dispersion persists in the extreme precipitation range (~>60 mm $d^{-1}$), with conservative estimated upper bounds (rarely exceeding 60 mm $d^{-1}$), suggesting that torrential events remain constrained by sample scarcity and feature-space boundary limitations. RF-Full amplifies mean bias from 0.478 to 1.282 mm $d^{-1}$, indicating that the model reduces overall error at the cost of increased mean offset—a trade-off between different error metrics.

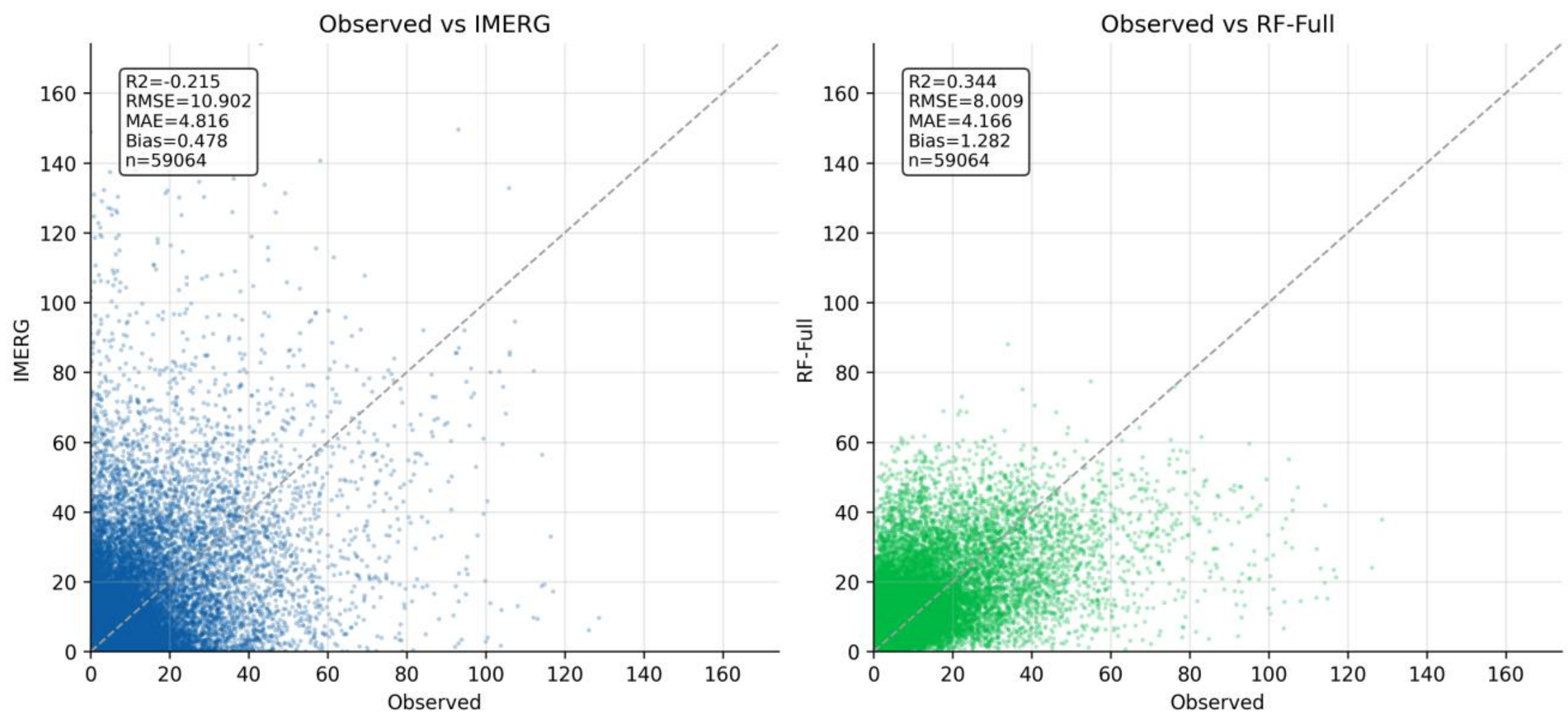

**Fig. 1.** Observed vs. estimated daily precipitation in Hunan Province, summer 2021–2022 (n = 59,064). (a) Raw IMERG ($R^2$ = −0.215, RMSE = 10.902 mm $d^{-1}$, MAE = 4.816 mm $d^{-1}$, Bias = 0.478 mm $d^{-1}$); (b) RF-Full ($R^2$ = 0.344, RMSE = 8.009 mm $d^{-1}$, MAE = 4.166 mm $d^{-1}$, Bias = 1.282 mm $d^{-1}$). Dashed line: 1:1 reference.

### 3.1.2. Numerical narrative of ablation experiments

Ablation experiments decompose the correction gain into incremental contributions from each predictor group (Table 1). The six-step progression from Raw IMERG to RF-Full reveals a steeply diminishing hierarchical structure: IMERG alone recovers $R^2$ to 0.066 (+0.281), indicating that the raw product error is structural; relative to this IMERG-only baseline, adding DEM alone yields +0.062, whereas adding the ERA5 atmospheric variables alone yields +0.225—a 3.6-fold asymmetry that indicates the dominant physical control of moisture and circulation (steps 3 and 4 are alternative, non-nested configurations, so both gains are measured against the IMERG-only baseline; because sequential ablation attributes shared variance to the predictor entered first, this ordering should still be read as indicative rather than exact). Once the linear combination (LR-Full) reaches $R^2$ = 0.344, RF-Full adds merely +0.001, with RMSE nearly unchanged (8.015 vs. 8.009). This near-additivity leaves little exploitable nonlinearity for tree-based models under the present feature set and sample size. The sole quantitative advantage of RF lies in a 14% MAE reduction (4.818 → 4.166), attributable to improved tail fitting for moderate rainfall, yet at the cost of inflating the mean bias from 0.454 in LR-Full to 1.282 mm $d^{-1}$ in RF-Full (raw IMERG: 0.478)—errors are redistributed from pointwise residuals toward the distributional mean. The MAE of LR-Full (4.818 mm $d^{-1}$) slightly exceeds that of the IMERG+ERA5 configuration (4.359 mm $d^{-1}$) and is comparable to raw IMERG (4.816 mm $d^{-1}$): because least-squares fitting prioritizes variance reduction over absolute error, individual metrics must be interpreted jointly rather than in isolation.

**Table 1.** Ablation experiment results for summer precipitation bias correction in Hunan Province. Models are ordered by increasing complexity. RF-Full marginally

exceeds LR-Full in $R^2$ (+0.001), indicating that the terrain–moisture relationship is predominantly additive, while the MAE reduction (−14%) reflects improved tail fitting. Full predictor set: IMERG + DEM + tcwv + u10 + v10. Metrics computed on the independent test period (2021–2022), n = 59,064.

| Model | $R^2$ | RMSE (mm $d^{-1}$) | MAE (mm $d^{-1}$) | Bias (mm $d^{-1}$) |
|---|---|---|---|---|
| Raw IMERG | −0.215 | 10.902 | 4.816 | 0.478 |
| IMERG | 0.066 | 9.558 | 4.925 | 0.672 |
| IMERG+DEM | 0.128 | 9.239 | 4.971 | 0.926 |
| IMERG+ERA5 | 0.291 | 8.332 | 4.359 | 1.231 |
| LR-Full | 0.3435 | 8.015 | 4.818 | 0.454 |
| RF-Full | 0.3445 | 8.009 | 4.166 | 1.282 |

The ablation experiments thus reveal that the terrain–moisture relationship is predominantly additive at the domain-aggregate scale: the linear formulation already captures the bulk of the domain-wide statistical structure. Yet the scatterplot diagnosis complicates this picture—although RF-Full pulls the overall scatter cloud toward the 1:1 line, significant dispersion persists in the extreme-precipitation zone, and the tail-fitting gain comes with amplified mean bias. A tension thus exists between global additivity and local validity: the linear model is "good enough" in an average sense, yet the RF improvement concentrated in moderate-to-heavy precipitation tails indicates that key mechanistic deviations persist within local subsets. The following sections diagnose these conditions through SHAP-based analysis across the TMI space.

The divergence between $R^2$ and MAE has a coherent metric-level interpretation. $R^2$ weights squared errors and is therefore dominated by rare high-magnitude events, whereas MAE weights all errors equally. Tree-based models cannot extrapolate beyond the training target range: their predictions are weighted averages of observed values, so extreme values are inevitably shrunk toward the conditional mean. RF thus leaves the extreme-driven squared-error budget essentially unchanged ($R^2$ +0.001) while reducing typical absolute errors in the data-dense light-to-moderate range (MAE −14%); the amplified mean bias is the asymmetric cost of this shrinkage. What RF learns beyond LR is therefore local conditional-mean refinement where data are abundant, not new large-scale structure, which the linear formulation already captures.

### *3.2. Mechanistic interpretation: TMI three-dimensional SHAP diagnosis*

#### 3.2.1. Global SHAP summary: Variable hierarchy and directional patterns

The SHAP results reveal that RF-Full responds to the major predictors with a clear, interpretable structure (Fig. 2a). The global importance ranking follows imerg > tcwv > v10 > dem > u10, yet the directional information carries greater explanatory value than the ranking alone. IMERG exhibits the widest SHAP distribution range, indicating that the model still relies primarily on the original satellite precipitation signal; low imerg values (blue) cluster in the negative SHAP region, whereas high imerg values (red) skew toward positive SHAP, reflecting a conditional co-directional adjustment to the raw signal. TCWV displays a pronounced piecewise nonlinear pattern: contributions are weak or negative under low moisture conditions, while positive contributions intensify under high moisture. This pattern is further manifested in the dependence plot as an accelerating upward relationship with increasing tcwv (Fig. 2b), and grid cells at high elevation (>1000 m) show stronger positive SHAP under equivalent tcwv, suggesting that terrain–moisture synergy serves as an important modulating factor of correction magnitude. V10 exhibits a predominantly negative offset trend: higher meridional wind components more frequently correspond to negative SHAP, indicating a consistent modulating effect of the wind field background on model output. By contrast, DEM shows bidirectional effects of moderate magnitude, and u10 contributions cluster near zero with overall weak influence.

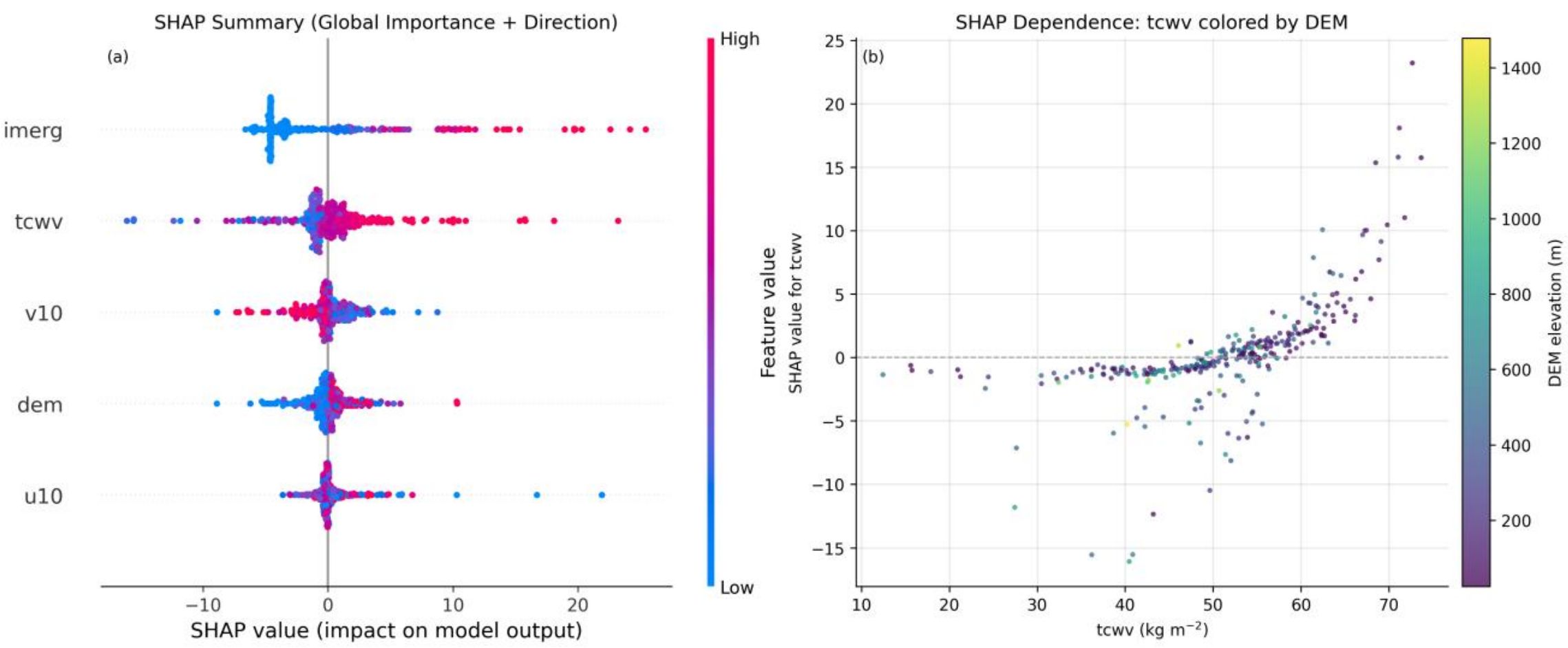

**Fig. 2.** SHAP explanation for the RF-Full (summer) model. (a) Summary beeswarm plot showing variable importance and directional effects. (b) SHAP dependence plot for tcwv, colored by DEM elevation (m) at 0.25° grid resolution, illustrating terrain–moisture synergy: positive contributions increase under high tcwv, and this gain is more pronounced in elevated terrain.

### 3.2.2. Terrain–moisture synergy: The tcwv–DEM dependence plot

The dependence plot (Fig. 2b) reveals the nonlinear coupling between moisture supply and topographic elevation that underpins the TMI framework. The tcwv contribution remains near zero below approximately 40 kg $m^{-2}$, but rises sharply above 50 kg $m^{-2}$, marking the activation of correction mechanisms under moisture-sufficient conditions. The color dimension—DEM elevation—uncovers a critical interaction: low-elevation points (purple, <400 m) remain near the lower envelope of the SHAP distribution even under high tcwv, whereas mid-to-high elevation points (green/yellow, >800 m) tend to occupy the upper tail, indicating that terrain uplift amplifies the positive gain of moisture correction. This elevation-conditioned gradient confirms that orographic modulation does not operate independently, but rather depends on moisture supply: terrain mechanisms become effective only when atmospheric moisture exceeds a necessary threshold.

Above 60 kg $m^{-2}$, SHAP values exhibit marked dispersion (with relatively sparse samples), suggesting that the model applies stronger positive corrections under high-moisture, high-terrain conditions, possibly reflecting a compensatory tendency for IMERG's underestimation of intense precipitation, yet the stability of the learned relationship declines under these extreme conditions. The 0.25° DEM resolution (approximately 25–28 km) captures broad-scale terrain contrasts—western mountains versus northern plains—but cannot resolve ridge-scale features; consequently, the tcwv–DEM interaction reflects areal-scale orographic modulation rather than fine-scale triggering. This resolution constraint sets the achievable upper bound of spatial correction performance for the current model.

### 3.2.3. Intensity-stratified SHAP dynamics: Directional inversion and mechanism degradation

SHAP analysis across the precipitation intensity spectrum reveals systematic mechanistic degradation toward extreme rainfall (Table 2). imerg maintains Spearman ρ above 0.95 across all classes, and tcwv above 0.80 (above 0.91 in all but the light-rain class), confirming their role as robust predictors largely invariant to intensity changes. dem ρ decays from 0.773 (light rain) to 0.676 (torrential rain), indicating attenuated topographic modulation as terrain-triggered convection transitions to large-scale organized systems.

The most diagnostically significant trend is the weakening of v10: its absolute value attenuates from −0.782 (light rain) to −0.356 (torrential rain), a reduction of approximately 54%. This marks the failure of meridional wind modulation logic—under moderate intensities, strong southerly winds associate with systematic negative corrections, but this relationship collapses under extreme intensities when convective organization dominates. The model faces a mechanistic dilemma: the same southerly wind that requires downward correction in moderate rainfall becomes a driver of intense precipitation in torrential rain, rendering the learned negative association invalid.

The contribution of u10 reverses polarity across rain-rate tiers: it remains positive from light to heavy rain (ρ = +0.096, +0.126, and +0.066, respectively) but flips to −0.207 in torrential rain. This reversal indicates that the physical role of zonal wind shifts across the convective spectrum. Under light-to-heavy intensities, positive u10 SHAP likely reflects the modulation of westerly components on moisture convergence, whereas under torrential intensities u10 no longer carries an identifiable precipitation-controlling signal, and the correction direction assigned by the model loses its physical basis. This reversal serves as a warning indicator of mechanistic failure: when a variable's SHAP direction flips, the learned relationship may have been extrapolated beyond the regime in which it was identified. However, u10 is the globally least important predictor and these correlations are small in magnitude; the reversal is therefore interpreted as a qualitative warning signal whose uncertainty is not formally quantified.

**Table 2.** Spearman correlation coefficients (ρ) for SHAP directional consistency by precipitation intensity class. Positive values indicate same-direction association between feature value and SHAP contribution; negative values indicate opposite-direction association. Direction denotes the dominant sign across classes; u10 reverses to negative at torrential intensity (−0.207; Section 3.2.3).

| Feature | Light SHAP ρ | Moderate SHAP ρ | Heavy SHAP ρ | Torrential SHAP ρ | Direction |
|---|---|---|---|---|---|
| imerg | 0.953 | 0.969 | 0.970 | 0.971 | + |
| tcwv | 0.806 | 0.916 | 0.920 | 0.917 | + |
| dem | 0.773 | 0.763 | 0.726 | 0.676 | + |
| v10 | −0.782 | −0.639 | −0.527 | −0.356 | − |
| u10 | 0.096 | 0.126 | 0.066 | −0.207 | +→− |

#### 3.2.4. Formalizing the TMI framework: From empirical patterns to diagnostic dimensions

The intensity-stratified dynamics in Table 2 naturally abstract into three orthogonal diagnostic dimensions—the Terrain–Moisture–Intensity (TMI) framework. Terrain (DEM) provides stable spatial anchoring: its SHAP direction remains positive across all intensity classes, but its magnitude attenuates toward torrential rain, thereby defining the boundary of topographic modulation saturation. Moisture (tcwv) serves as the dynamic principal driver: its SHAP magnitude increases systematically with precipitation intensity and determines whether terrain correction activates. Intensity governs the failure mode: the directional reversal of u10 and attenuation of v10 mark the threshold beyond which the learned terrain–moisture synergy loses validity.

The unifying concept is mechanism purity (Section 1): under high-purity conditions (coherent terrain, stable season, moderate intensity), the additive terrain–moisture relationship holds and correction succeeds; under degraded purity (mixed terrain, transitional season, extreme intensity), variables shift from synergistic predictors to conflicting signals, and correction fails. Thus, the three TMI boundaries—spatial, temporal, and intensity—are not independent constraints, but different manifestations of the same underlying principle: machine learning-based correction is valid only within the range where the training mechanism retains its purity.

### *3.3. Spatial boundary: Regional heterogeneity of terrain coupling*

#### 3.3.1. DEM–SHAP Curves: Four Terrain Response Patterns

Subregional comparison of DEM–SHAP relationships reveals pronounced spatial heterogeneity (Fig. 3). North Hunan (green, n = 4,911), situated in the plain–hill transition zone, exhibits the steepest positive rise at low elevations (0–200 m; SHAP

−2 to +2), suggesting that modest topographic relief generates significant correction signals. West Hunan (blue, n = 3,706) maintains sustained amplification in the mid-elevation range (200–800 m), approaching a plateau above 800 m (SHAP ≈ +3) with diminishing marginal effects. South Hunan (orange, n = 4,030) shows a gradual rise across 200–1,400 m with the shallowest overall slope, indicating stable but relatively weak terrain dominance. Central Hunan (red, n = 2,353) presents a nearly flat curve with scatter widely distributed around the zero line, indicating the weakest consistency in DEM contributions—consistent with mechanistic overfitting, wherein variables are activated yet physical relationships lack stability. Overall, DEM contributions are positive across all subregions, but response intensity and saturation characteristics vary markedly.

To control scatter density, Fig. 3 displays a random subsample of the test set (n = 15,000, drawn without replacement, random_state = 42); the n values in parentheses denote counts within this subsample, not the full-sample partition sizes reported in Table 3.

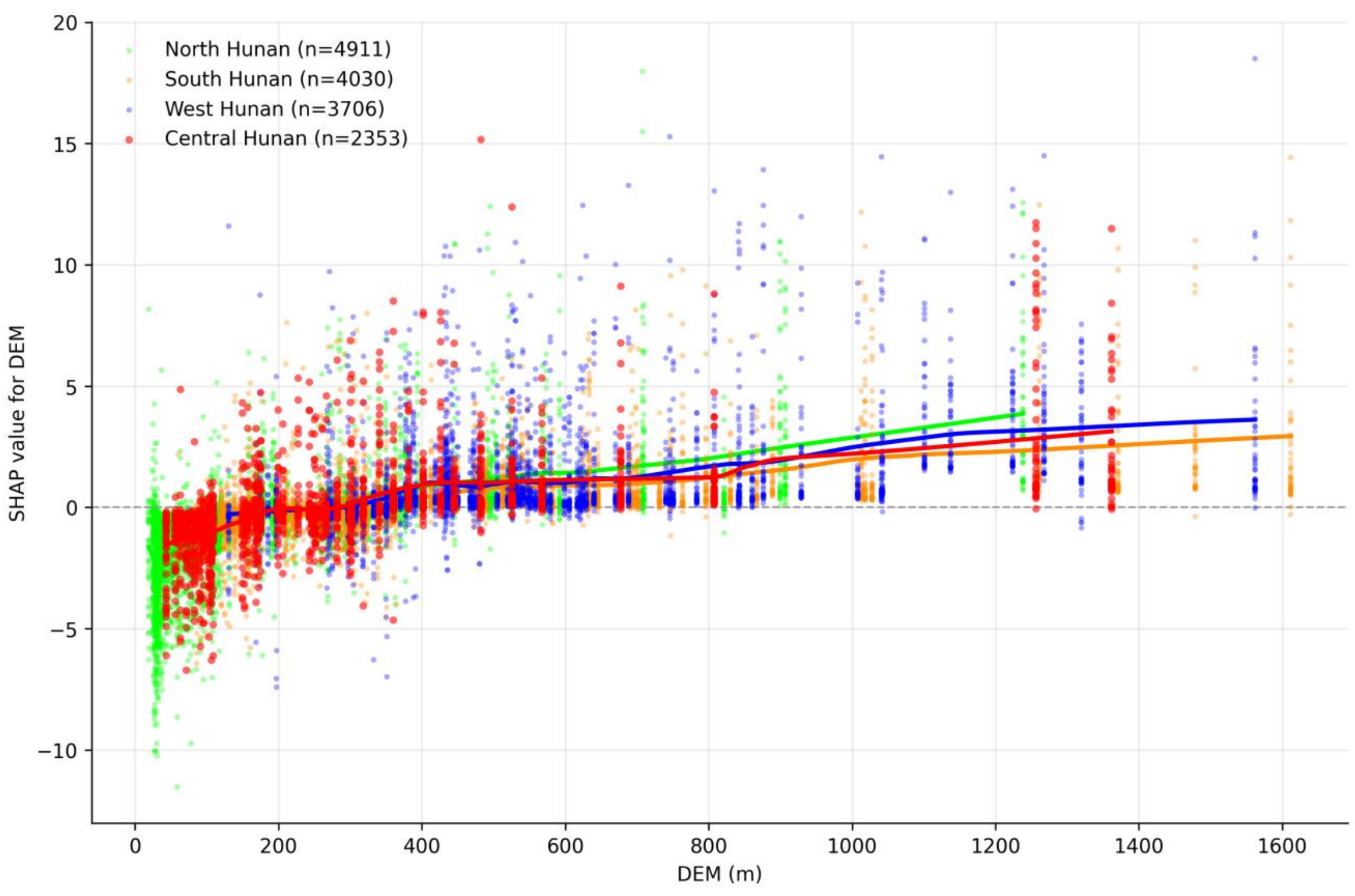

**Fig. 3.** DEM–SHAP scatter and fitted curves for four subregions (random subsample of the test set, n = 15,000, drawn without replacement, random_state = 42; legend n denotes per-subregion subsample counts; full-sample sizes in Table 3). North Hunan (green, n = 4,911) shows the steepest low-elevation slope; West Hunan (blue, n = 3,706) rises through mid-elevations, plateauing above 800 m; South Hunan (orange, n = 4,030) rises gradually; Central Hunan (red, n = 2,353) is flat with scatter near zero.

### 3.3.2. Performance–SHAP Paradox: The Central Hunan Collapse

The subregional performance gradient (Table 3) confirms that correction skill is governed by terrain mechanism purity, not elevation. South Hunan achieves the highest $R^2$ (0.453) and lowest RMSE (6.967 mm $d^{-1}$), whereas Central Hunan collapses to $R^2 = 0.133$ with bias inflating to 2.161 mm $d^{-1}$. The $R^2$ gap between Central and South Hunan is statistically significant (bootstrap $\Delta R^2 = -0.320$, 95% CI [−0.404, −0.237], $p < 0.001$), confirming that the performance difference is not sampling noise.

The paradox lies in Central Hunan's relatively high univariate SHAP directional consistency (tcwv $\rho = 0.875$, v10 $\rho = -0.763$) coexisting with the lowest predictive skill. Central Hunan's RMSE (7.243 mm $d^{-1}$) is lower than West Hunan (9.003) or North Hunan (8.369), indicating small error magnitude yet systematic directional bias. This pattern—directional error under low variance—defines what we term mechanistic overfitting, a diagnostic notion distinct from statistical overfitting: rather than fitting sampling noise in the training period (which would manifest as a train–test performance gap), the model over-applies terrain-correction logic learned in coherent-terrain regimes to stratiform-dominated grid cells, generating systematic positive bias through erroneous upward adjustment. Because $R^2$ is normalized by observational variance, part of the subregional $R^2$ contrast inevitably reflects differences in precipitation variance: back-calculating variance from $R^2$ and RMSE in Table 3 yields approximately 60 (Central), 89 (South), 94 (North), and 135 (West) mm$^2$ $d^{-2}$, so the $R^2$ ordering partly mirrors variance structure. Two observations, however, indicate that variance compression alone cannot account for Central Hunan's collapse: its mean bias (2.161 mm $d^{-1}$) is 3.5 times that of South Hunan (0.614 mm $d^{-1}$) despite its lower variance, and it combines the lowest RMSE with the largest systematic bias—a directional failure signature rather than a variance artifact.

Data-driven K-means clustering identifies a cluster that concentrates high-precipitation grid-days (mean observed precipitation 29.2 mm $d^{-1}$, versus 2.8–3.0 mm $d^{-1}$ in the three geographic clusters), whose geographic centroid falls within the Central Hunan region (Supplementary Table S2). This finding is consistent with the subjective partition based on topographic coherence. Because the clustering features include geographic coordinates and precipitation-related variables, this analysis is used as a consistency check rather than fully independent validation of the subregional division.

This high-precipitation regime implies a distinct variance structure; however, regime difficulty alone cannot explain why near-maximal variable activation yields the lowest predictive skill (Table 3)—strong SHAP engagement that fails to convert into skill is a signature of conflicting coexisting mechanisms rather than of an inherently harder prediction problem.

**Table 3.** Subregional performance and SHAP directional consistency for RF-Full (n_estimators = 500). South Hunan achieves the highest $R^2$ (0.453) owing to coherent terrain; Central Hunan collapses ($R^2 = 0.133$) owing to mechanism purity

degradation from mixed terrain. The spatial boundary is defined by terrain mechanism purity, not elevation.

| Subregion | DEM mean (m) | N | $R^2$ RF | RMSE RF | Bias RF | imerg ρ | tcwv ρ | dem ρ | v10 ρ |
|---|---|---|---|---|---|---|---|---|---|
| West Hunan | 569 | 14,720 | 0.401 | 9.003 | 0.812 | 0.951 | 0.661 | 0.523 | −0.743 |
| Central Hunan | 282 | 9,200 | 0.133 | 7.243 | 2.161 | 0.937 | 0.875 | 0.706 | −0.763 |
| North Hunan | 203 | 18,952 | 0.254 | 8.369 | 1.791 | 0.960 | 0.898 | 0.700 | −0.744 |
| South Hunan | 432 | 16,192 | 0.453 | 6.967 | 0.614 | 0.948 | 0.739 | 0.723 | −0.747 |

### 3.3.3. Mechanism Purity Ranking and Spatial Boundary

Four subregions correspond to four distinct terrain–precipitation coupling regimes, ordered by decreasing mechanism purity (Table 3). South Hunan represents the canonical high-purity terrain type: the coherent Nanling Mountains provide a clear orographic face for the summer monsoon, with DEM, tcwv, and moisture advection physically aligned, yielding stable and strong correction. West Hunan also exhibits high purity, but with saturation constraints—skill is high yet bounded by an ~800 m elevation ceiling, beyond which marginal returns diminish. North Hunan is stratiform-dominated: flat plains eliminate terrain complexity, reducing correction to moisture–large-scale dynamics and producing moderate but directionally stable skill. Central Hunan represents the lowest-purity type: mixed hill–plain mosaic fragments precipitation signals, forcing the model to confuse terrain and stratiform logic, generating maximum variable activation with minimum predictive skill.

The physical root of Central Hunan's collapse lies in terrain fragmentation. The area-averaged 0.25° DEM (~28 km) represents the hill–basin mosaic as a single elevation value (282 m), unable to resolve intra-grid terrain variation. Located on the southeastern flank of the Xuefeng Mountains, Central Hunan's complex topography mixes lee-slope dynamics with plain stratiform processes, degrading terrain–moisture coupling consistency. Elevation height shows no clear correlation with correction skill (North Hunan, 203 m, $R^2$ = 0.254 vs. South Hunan, 432 m, $R^2$ = 0.453—a nearly two-fold difference—and a more than three-fold gap between Central and South Hunan), indicating that the TMI spatial boundary is a purity threshold rather than an elevation threshold.

### 3.3.4. A priori predictability of the spatial boundary

The a priori coherence proxy identifies Central Hunan as the least organized precipitation regime before any model is consulted: mean daily Moran's I of observed precipitation is 0.555 in Central Hunan, versus 0.620 in South Hunan, 0.686 in North Hunan, and 0.689 in West Hunan (Supplementary Table S5). The fragmented spatial structure of daily precipitation in Central Hunan is consistent with the coexistence of incompatible mechanisms (lee-slope dynamics and plain stratiform processes) posited in Section 3.3.2 and constitutes a model-independent signature of degraded mechanism purity.

Because $R^2$ is normalized by observational variance (Section 4.4), we additionally evaluate correction efficiency as the percentage RMSE reduction relative to raw IMERG, a metric independent of variance normalization. Efficiency reaches 26–30% in the three coherent regions (South 26.6%, West 30.1%, North 26.2%) but only 16.2% in Central Hunan (Supplementary Table S5). The variance-free metric thus confirms that Central Hunan's failure is not an artifact of its low observational variance (60 $mm^2\ d^{-2}$, the lowest among the four subregions), although that variance amplifies the $R^2$ contrast. Together, these results support a two-factor interpretation: mechanism purity gates learnability—below a coherence threshold, correction efficiency collapses—whereas observational variance sets the $R^2$ scale among learnable regions. The coherence proxy is computable from historical observations alone, demonstrating that mechanism purity can predict, rather than merely describe post-hoc, where correction will fail. At the resolution supported by both metrics, the agreement is ordinal rather than exact: Central Hunan is separated from the three coherent subregions by more than three standard errors in the coherence proxy and by at least ten percentage points in efficiency, whereas differences among the three coherent subregions lie within their respective uncertainties on both axes. The framework's a priori prediction is thus the identification of the low-purity regime rather than a fine-grained ranking, and that prediction is confirmed.

## *3.4. Temporal boundary: Seasonal differences and mechanism mismatch*

### 3.4.1. Daily bias evolution: Distinct mechanistic phases

The 5-day moving average of daily mean bias reveals phase-dependent characteristics associated with the Mei-yu to post-Mei-yu transition (Fig. 4). During the Mei-yu onset (early June), both 2021 and 2022 exhibit large fluctuations (approximately −4 to +5 mm $d^{-1}$), reflecting the instability of frontal processes prior to the full organization of the monsoon trough. During the Mei-yu peak (mid-June to early July), the two years differ in bias magnitude: RF bias in 2021 is mostly +2 to +4 mm $d^{-1}$, whereas in 2022 it is mostly +1 to +2 mm $d^{-1}$, suggesting that interannual background conditions modulate correction amplitude. This indicates that the Mei-yu correction response does not correspond to a single fixed polarity, but adjusts with synoptic conditions.

In the post-Mei-yu period (mid-July to August), terrain convection dominates and IMERG microwave undersampling is systematic, with both years showing predominantly positive bias. In 2022, the bias remains stable at +1 to +2 mm $d^{-1}$. In 2021, the same positive bias prevails, but a significant spike exceeding +9 mm $d^{-1}$ occurs in early August, suggesting that under extreme convective conditions, model compensation may approach or exceed the coverage of the training sample, with risk of mechanistic failure. The convergence of IMERG and RF trajectories in early June of both years marks the transition boundary: during the early stage of seasonal transition, the mechanistic constraints available for stable correction are relatively weak.

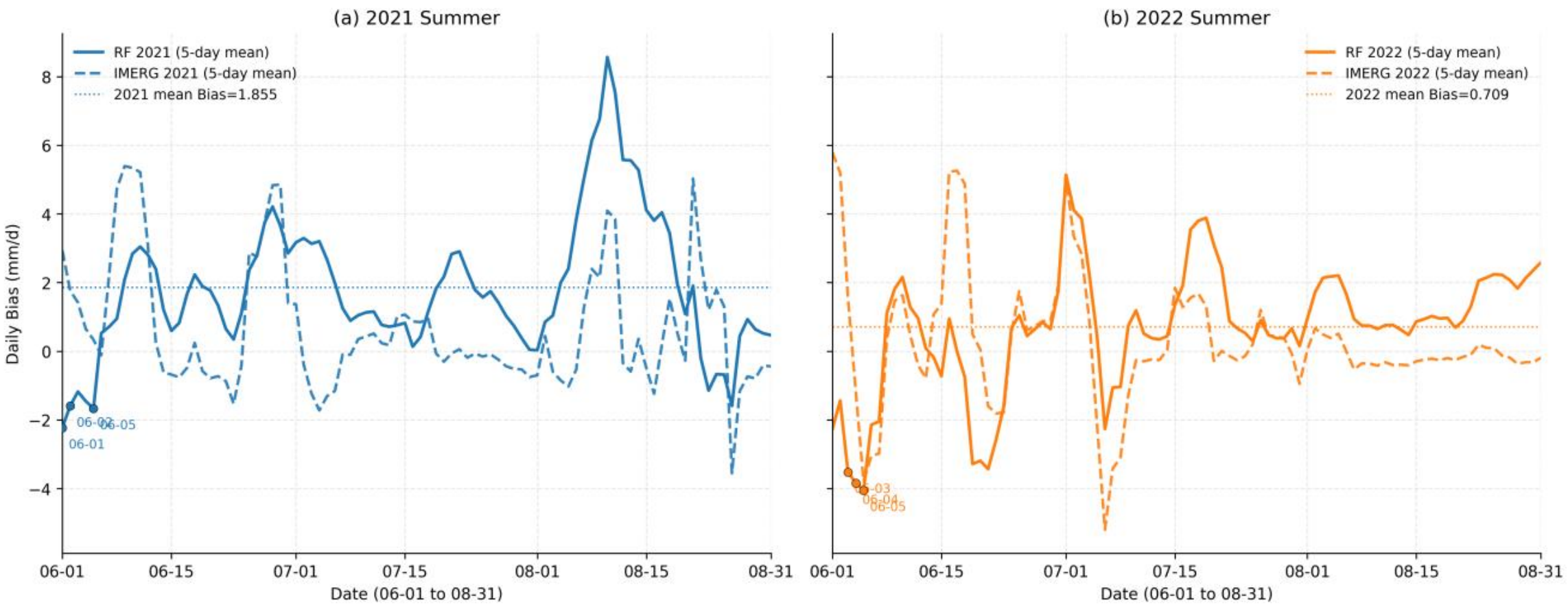

**Fig. 4.** Five-day moving average of daily mean bias in summer (June 1-August 31) for (a) 2021 and (b) 2022. Solid line: RF-Summer; dashed line: raw IMERG; dotted horizontal line: seasonal mean bias for each year. Bias is defined as estimate minus observation; positive values indicate overestimation. A prominent positive spike appears in early August 2021 (~+9 mm $d^{-1}$), whereas 2022 remains mostly around +1 to +2 mm $d^{-1}$.

### 3.4.2. Interannual performance: Regime-dependent algorithmic advantage

Interannual comparison is presented in Table 4. In 2021, characterized by more persistent Mei-yu processes and a higher proportion of stratiform precipitation, IMERG exhibited negative skill ($R^2$ = −0.390), with linear regression marginally outperforming random forest (LR $R^2$ = 0.292 vs. RF $R^2$ = 0.275, $\Delta R^2$ = −0.017). In 2022, with more frequent convective transitions, RF-Full achieved $R^2$ = 0.406, exceeding LR-Full ($\Delta R^2$ = +0.017). This reversal indicates that the relative model advantage exhibits marked precipitation-regime dependence, rather than being determined solely by model formulation.

Bias trajectories further support this interpretation: RF-Full bias in 2021 reached 1.855 mm $d^{-1}$, substantially exceeding the IMERG bias for that year (0.655 mm $d^{-1}$), whereas in 2022 it decreased to 0.709 mm $d^{-1}$. Meanwhile, LR-Full in 2022 approached near-neutral bias (Bias = 0.061 mm $d^{-1}$), suggesting that linear correction also achieved reasonable adaptation under those conditions. The pooled metrics ($R^2$ = 0.344, Table 1) are obtained through cross-year aggregation, which may attenuate interannual mechanistic differences.

**Table 4.** Interannual performance comparison (2021 vs. 2022). RF-LR $\Delta R^2$ reverses from −0.017 to +0.017. RF bias amplifies to 1.855 mm $d^{-1}$ in 2021 but decreases to 0.709 mm $d^{-1}$ in 2022. Pooled metrics in Table 1.

| Year | N | IMERG $R^2$ | IMERG Bias | LR $R^2$ | LR Bias | RF $R^2$ | RF Bias |
|---|---|---|---|---|---|---|---|
| 2021 | 29,532 | −0.390 | 0.655 | 0.292 | 0.846 | 0.275 | 1.855 |
| 2022 | 29,532 | −0.059 | 0.302 | 0.389 | 0.061 | 0.406 | 0.709 |

### 3.4.3. Seasonal SHAP triptych: Visual mechanism comparison

Figure 5 presents three SHAP configurations: (a) RF-Summer applied to spring; (b) RF-Summer applied to summer; (c) RF-Spring applied to spring. Overall, imerg and tcwv maintain positive contributions across all three configurations with relatively limited amplitude variation, indicating that the base precipitation signal and moisture-related correction exhibit good directional stability under cross-seasonal conditions.

Differences emerge primarily in wind-field variables: v10 shows strong negative contribution in summer (−0.744), weakening to negative values in spring configurations (−0.412 / −0.423); u10 shifts from weak positive contribution in summer (+0.096) to weak negative contribution in spring (−0.156 / −0.148). The v10 direction in panel (c) for RF-Spring (−0.423) closely matches the RF-Summer projection to spring (−0.412), suggesting that the meridional wind modulation direction in spring possesses some degree of portability, while its intensity varies seasonally.

Collectively, imerg, tcwv, dem, and v10 demonstrate overall directional stability in cross-seasonal migration, whereas u10 is the sole variable exhibiting directional reversal, indicating greater sensitivity to seasonal background and possible association with precipitation-regime differences between the training and target domains.

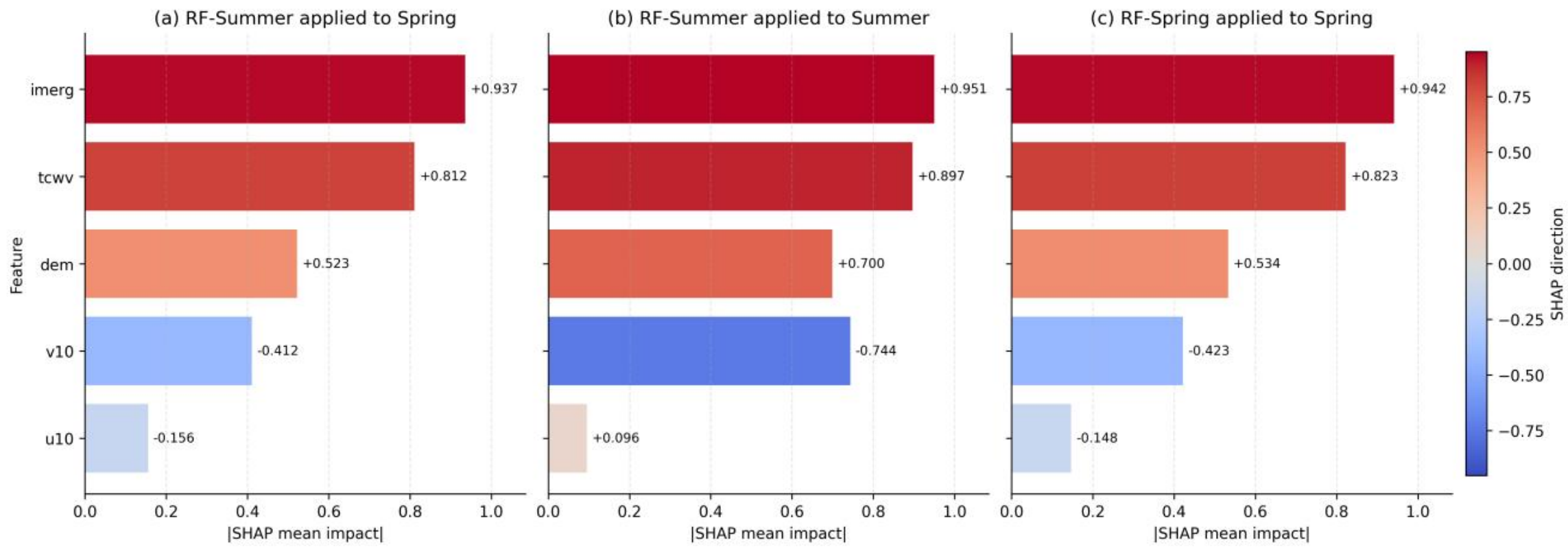

**Fig. 5.** Cross-seasonal SHAP directional comparison. (a) RF-Summer applied to spring; (b) RF-Summer applied to summer; (c) RF-Spring applied to spring. imerg and tcwv maintain positive contributions across configurations; v10 weakens in spring; u10 reverses sign (Table 5).

### 3.4.4. Directional decomposition: Single-variable sensitivity

u10 is the primary exception variable (Table 5): weakly positive in the summer-native model, yet negative in both the cross-seasonal projection and the spring-native model. Because the two spring-domain values are close (difference 0.008), this directional reversal is unlikely to arise solely from migration noise, and more likely reflects changes in wind-field dynamics under differing seasonal backgrounds.

This implies that the summer-trained model did not fully learn the spring polarity of u10; during cross-seasonal application, overall performance does not collapse abruptly, yet systematic deviation occurs at variables requiring directional correction—i.e., "silent failure." Spring possesses a stable, identifiable physical mechanism (RF-Spring→Spring achieves 5/5 directional consistency); the failure lies in RF-Summer's inability to learn the season-specific polarity of u10.

At the variable level, the temporal boundary of the TMI framework is primarily characterized by the polarity difference of u10, rather than comprehensive mechanism mismatch: four of five variables maintain stable directions, with only u10 reversing. This phenomenon echoes the RF-LR advantage reversal and bias polarity shifts in the interannual analysis, suggesting that correction portability depends on consistency in key mechanistic variables between the training and target domains.

**Table 5.** Seasonal SHAP directional comparison. Values are mean SHAP contributions (mm $d^{-1}$); the sign indicates the direction of the adjustment relative to the prediction baseline. Cross-Season indicates directional stability of RF-Summer across seasons; SpringDomain indicates consistency between RF-Summer and RF-Spring within the spring domain. u10 is the only variable that switches direction across seasons.

| Feature | Summer→Spring | Summer→ Summer | Spring→ Spring | Cross-Season | Spring Domain |
|---|---|---|---|---|---|
| imerg | 0.937 | 0.951 | 0.942 | Yes | Yes |
| tcwv | 0.812 | 0.897 | 0.823 | Yes | Yes |
| dem | 0.523 | 0.700 | 0.534 | Yes | Yes |
| v10 | −0.412 | −0.744 | −0.423 | Yes | Yes |
| u10 | −0.156 | 0.096 | −0.148 | No | Yes |

### *3.5. Intensity boundary: Progressive failure at extreme precipitation*

#### 3.5.1. Feature Space Continuity: Challenging the Domain-Shift Hypothesis

The t-SNE projection of the feature space (n = 822 stratified subsample) shows that torrential-rain events (≥50 mm $d^{-1}$, red, n = 522) do not form fully isolated clusters (Fig. 6a). Instead, heavy-precipitation samples extend continuously from the non-torrential subdomain (<50 mm $d^{-1}$, blue, n = 300): partial overlap exists in the central region (approximately −10 to +10 on both axes), while simultaneously expanding toward the high-tcwv–high-DEM region on the right. This continuity suggests that extreme precipitation is not a mechanistically independent "foreign class," but rather approaches the upper tail of an intensity continuum. A t-SNE perplexity sensitivity analysis (perplexity = 5, 10, 30, 50, 100) confirms that the partial-overlap pattern between torrential and non-torrential events is robust to perplexity choice (mean coefficient of variation across three stability metrics = 0.150; see Supplementary Fig. S1 and Table S3).

Using SHAP standard deviation (σ) to characterize mechanistic dispersion, an overall increasing trend across intensity classes is evident (Fig. 6b). IMERG exhibits the highest σ (approximately 9.8), showing marked increase from light precipitation; DEM displays the largest relative amplification. This combination of "spatial continuity with rising mechanistic dispersion" collectively indicates that the intensity boundary is defined by mechanistic saturation rather than by geometric isolation.

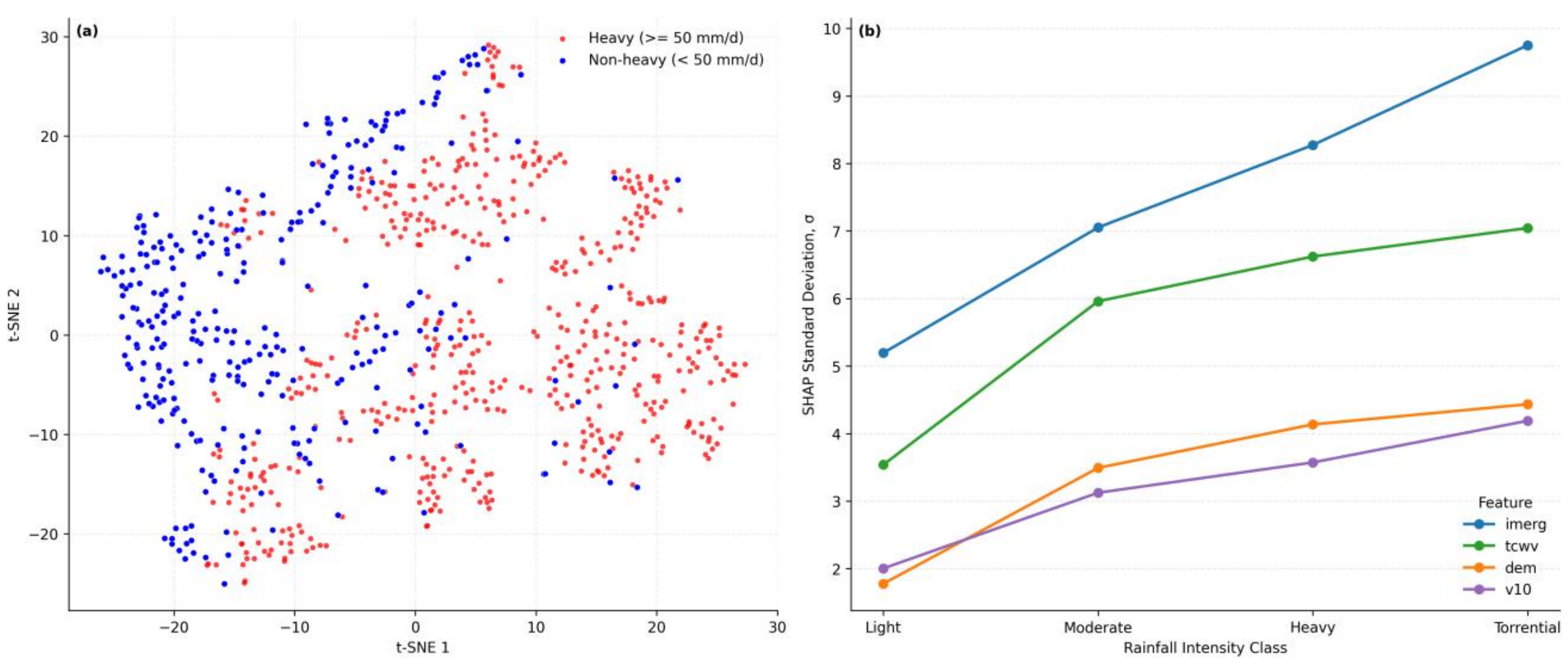

**Fig. 6.** Mechanistic decomposition under extreme precipitation intensity. (a) t-SNE projection of feature space (n = 822). Non-torrential samples (blue) partially overlap with torrential samples (red) in the central region, with torrential samples extending toward the high-tcwv–high-DEM region. (b) SHAP standard deviation (σ) across precipitation intensity classes.

### 3.5.2. SHAP disorder escalation: Quantifying mechanistic decay

Although extreme rainfall is spatially continuous, its correction logic is not. The SHAP disorder index—the standard deviation of SHAP values within each intensity class—quantifies this decay (Fig. 6b). All four predictors exhibit a monotonic increase in σ from light rain to torrential rain: imerg from ~5.2 to ~9.8 (approximately +88%), tcwv from ~3.5 to ~7.0 (+100%), v10 from ~1.8 to ~4.2 (approximately +133%), and dem from ~1.8 to ~4.5 (+150%). Bootstrap validation (N = 1,000) based on low-quantile (0–20%) versus high-quantile (80–100%) group comparisons confirms that σ differences for all four variables are significantly greater than zero (dem Δσ = 2.722 [2.656, 2.794], tcwv Δσ = 4.412 [4.300, 4.519], v10 Δσ = 1.893 [1.807, 1.970], imerg Δσ = 5.075 [4.899, 5.242]; all $p < 0.001$; see

Supplementary Table S4). Because SHAP values carry units of mm $d^{-1}$ and scale with the magnitude of the predicted correction, part of the σ increase is expected mechanically as intensity grows; we therefore interpret σ escalation as a relative, class-conditional dispersion diagnostic that complements—rather than substitutes for—normalized dispersion measures (Section 4.4).

The dem disorder surge is particularly diagnostic: stable terrain modulation at moderate intensities becomes the fastest-saturating component as orographic triggering yields to large-scale convective organization. Meanwhile, the v10 disorder escalation reflects fragmentation of the stratiform-correction logic. The coexistence of these trends indicates that the terrain–moisture synergy stabilizing moderate-rain correction progressively disintegrates at extreme intensities.

### 3.5.3. Conceptualizing the progressive failure boundary

Dual-panel evidence supports a mechanistic rather than spatial conceptualization of the intensity boundary. Three processes drive progressive failure. First, the tcwv–DEM coupling anchoring moderate-rain correction weakens: high humidity no longer guarantees predictable terrain response, as convective organization decouples precipitation from local terrain control. Second, dynamic variables (v10) transition from structured predictors to noise—their stratiform-correction logic fragments. Third, SHAP disorder itself serves as a real-time diagnostic: monotonic σ escalation is not merely a symptom of failure but a quantitative indicator of its progression.

This contrasts with traditional binary views—where correction is either effective or ineffective. Instead, the intensity boundary is a saturation frontier: as the terrain–moisture manifold extends toward its extrapolation limit, correction quality gradually degrades. At 50 mm $d^{-1}$ as an operational boundary for heavy precipitation, mechanistic dispersion increases markedly at the high-intensity end, and operational reliability declines significantly. Given the limited torrential sample, we present this saturation-frontier interpretation as a proof-of-concept diagnosis at the domain-aggregate level.

### 3.5.4. Unifying the three TMI boundaries

The three boundaries—spatial (Section 3.3), temporal (Section 3.4), and intensity (Section 3.5)—collectively support the interpretive framework that "mechanism consistency constrains correction portability." Spatial consistency requires unified terrain–precipitation coupling (single geomorphic or stratiform-dominated type); temporal consistency requires stable circulation dynamics (season-specific wind–precipitation relationships); intensity consistency requires terrain–moisture synergy to operate within its linearizable range. Where any consistency dimension degrades—mixed terrain (Central Hunan), seasonally transitional circulation (spring u10 reversal), or extreme intensity (disorder saturation)—correction degrades significantly. The TMI framework provides an operable diagnostic system: each failure mode can be traced to a specific boundary violation, and each boundary can be quantified through the SHAP-based indicators presented herein.

### 3.6. Cross-regional validation: pre-registered purity screening in Guangxi and Guangdong

#### 3.6.1. Design and pre-registered predictions

The spatial boundary results of Section 3.3 rest on a single province. To test whether the coherence proxy generalizes, the analysis is extended to Guangxi and Guangdong, chosen to hold the broad South China summer monsoon climate approximately constant while varying terrain and precipitation regime: karst northwestern Guangxi represents an extreme of terrain fragmentation, whereas coastal Guangdong adds a typhoon-influenced intensity spectrum. Both provinces are processed with the identical pipeline (Section 2.2; 329 and 243 valid 0.25° grid cells, respectively, after the finite-value mask), the same 2016–2020 training and 2021–2022 testing split, and geomorphic subregional stratifications defined in Section 2.3.5. Before any Guangxi or Guangdong model was evaluated, the coherence–efficiency relationship was fitted on the four Hunan subregions alone (efficiency = 84.30 × Moran's I − 28.98) and applied to the Guangxi and Guangdong proxies; the five subregion predictions were locked in a UTC-timestamped record (Supplementary Table S8), constituting a pre-registered out-of-sample test.

Guangxi exhibits the highest precipitation coherence in the study: mean daily Moran's I of observed precipitation reaches 0.787 in the karst northwest and 0.773 in the southeastern hills, versus 0.555–0.689 across the Hunan subregions and 0.664–0.707 across Guangdong (Fig. 7 and Supplementary Fig. S2). The geomorphic intuition that karst fragmentation should degrade coherence is therefore not supported at the 0.25° daily scale: summer daily precipitation fields over karst Guangxi are spatially well organized despite the complexity of the underlying terrain. Against the pre-registered rule, four of the five efficiency classes are predicted correctly, and the mean absolute error across the five subregions is 2.6 percentage points (Supplementary Table S7); the karst northwest prediction is accurate to 0.5 points (predicted 37.4%, realized 37.9%). The single class miss is instructive rather than damaging: the two Guangdong subregions outside the northern mountains are over-predicted by 4–7 points because the proxy measures signal coherence, whereas realized efficiency also depends on the satellite baseline—IMERG is already comparatively accurate over coastal Guangdong, leaving less headroom for correction (Section 4.4).

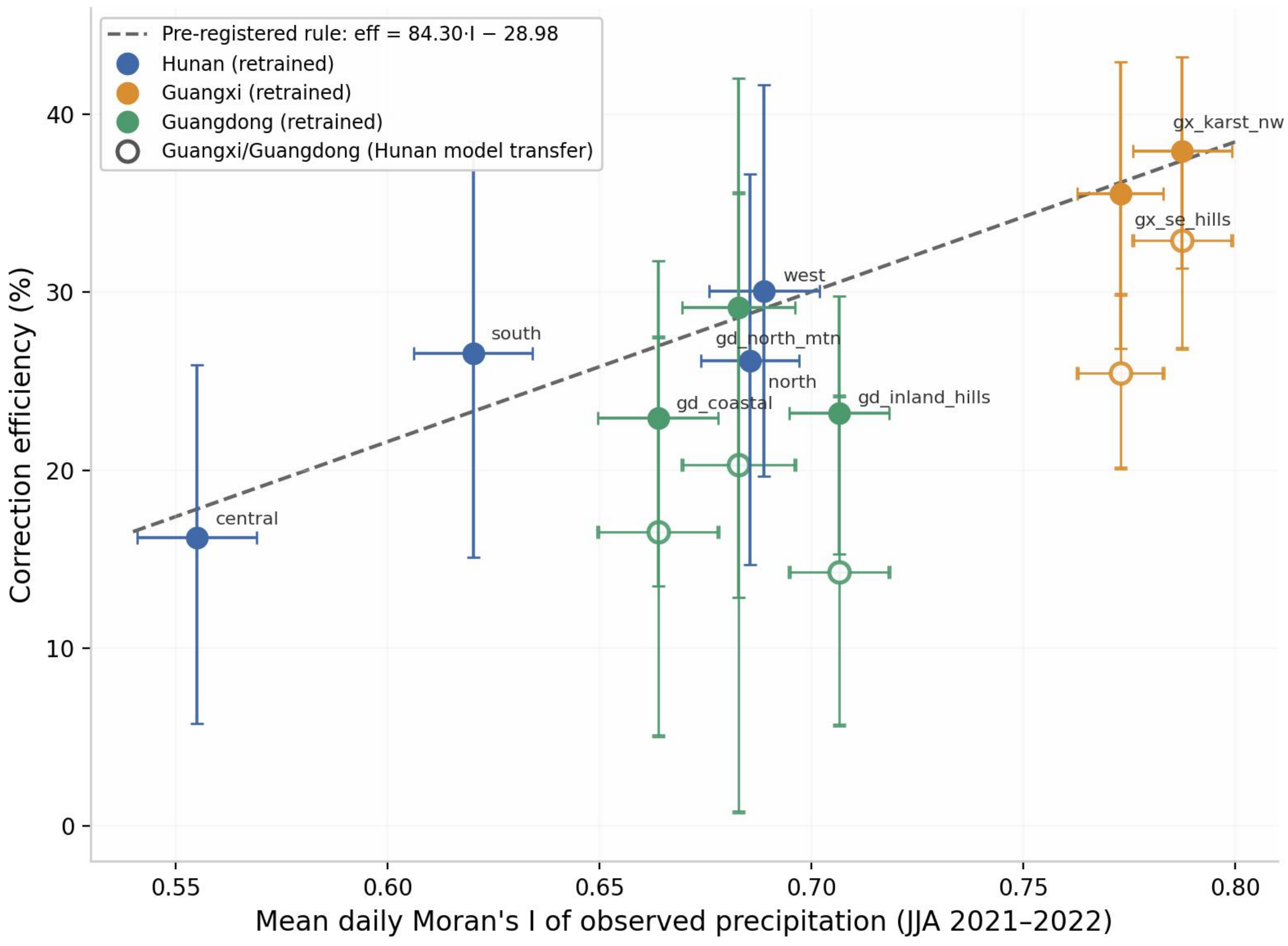

**Fig. 7.** Coherence proxy versus realized retrained RF correction efficiency for the nine subregions of Hunan, Guangxi, and Guangdong (JJA 2021–2022). The pre-registered rule (Table S8) is shown as the fitted line; error bars denote ±1 SE of Moran's I over days and day-block bootstrap 95% CIs of efficiency.

### 3.6.2. Two-arm test: mechanism mismatch versus portability

Direct transfer of the Hunan-trained model separates the two regimes sharply. Transferred to Guangxi, correction efficiency holds at 28.2%, comparable to its within-Hunan value (26.5%); transferred to Guangdong, it degrades to 16.9% (95% CI 8.4–25.1%). Local retraining reverses the pattern: Guangdong recovers to 25.0%, and Guangxi reaches 36.5%, the highest provincial efficiency in the study (Fig. 8 and Supplementary Table S6). The contrast between the arms isolates the mechanism effect: Guangdong's coherence is moderate and its local signal is learnable, but the predictor–predictand mapping learned in Hunan does not port to a typhoon-influenced coastal regime—mechanism mismatch, not learning failure. Notably, linear transfer outperforms random-forest transfer in both provinces (Guangxi 33.6% versus 28.2%; Guangdong 21.4% versus 16.9%), indicating that the forest partly overfits source-region mechanism structure, consistent with the near-additive terrain–moisture relationships identified in Section 3.1.

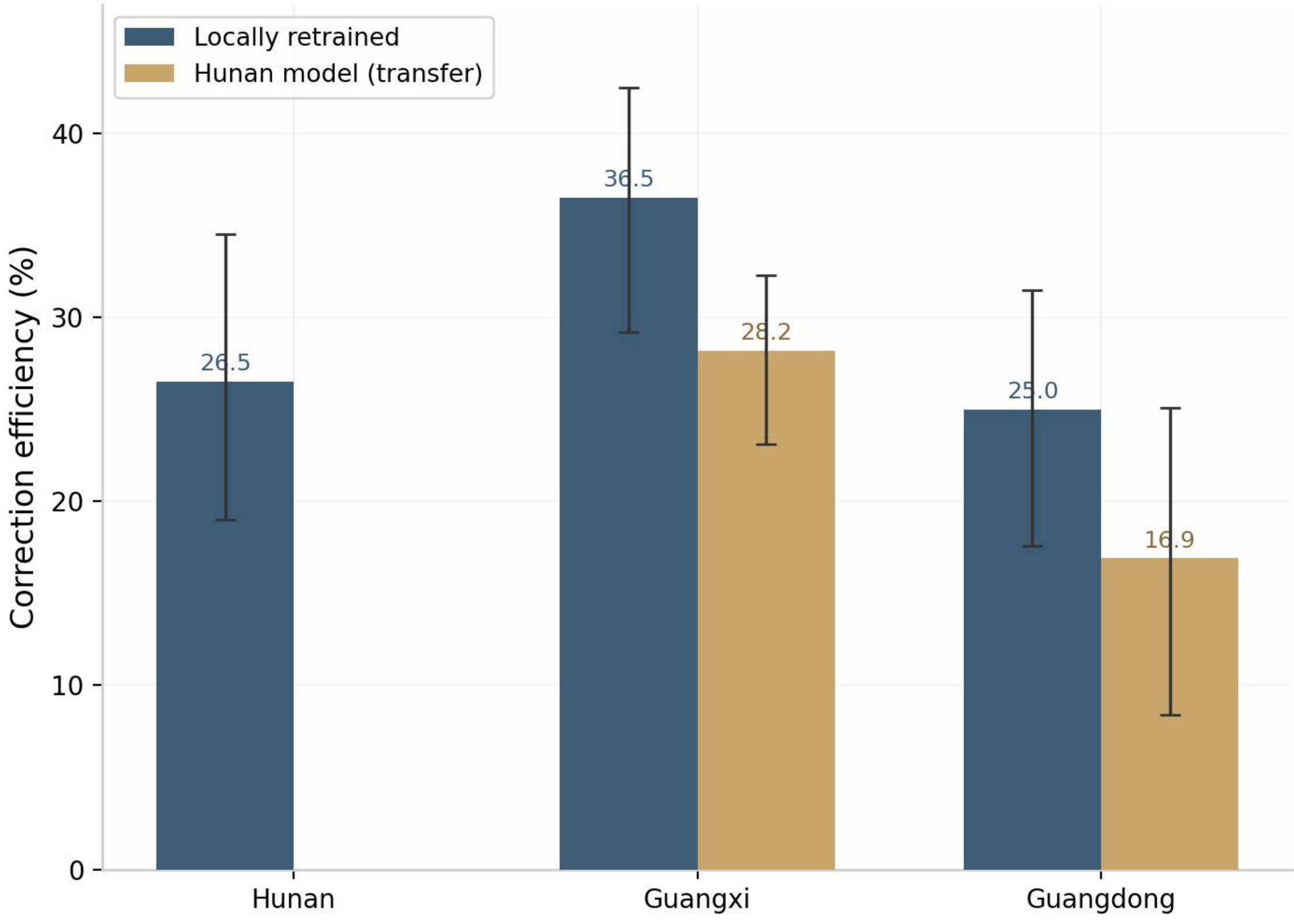

**Fig. 8.** Two-arm cross-regional comparison: transferred (Hunan-trained) versus locally retrained RF correction efficiency with day-block bootstrap 95% CIs for Guangxi and Guangdong; the single Hunan bar gives the within-region reference efficiency (26.5%). Values in Supplementary Table S6.

### 3.6.3. Leave-one-region-out consistency

Rotating source and target roles across the three provinces yields a fully crossed transfer matrix (Supplementary Table S6). Within-region training is always optimal: the diagonal efficiencies (Hunan 26.5%, Guangxi 36.5%, Guangdong 25.0%) exceed every off-diagonal entry in their column. Transfer between the two inland monsoon provinces retains most of the correction skill (Hunan→Guangxi 28.2%, Guangxi→Hunan 23.8%), whereas transfer into coastal Guangdong collapses to 16.9–19.0%, and Guangdong-sourced models lose 6–10 efficiency points relative to the target diagonal—the signature of mechanism mismatch, absent between regime-similar regions. At the provincial scale, mechanism purity additionally gates transferability: every province remains correctable by its own model—none approaches the low-coherence regime in which learnability collapses within Hunan (Section 3.3.4)—but portability holds only where the dominant precipitation mechanism is shared. Together with the pre-registered proxy predictions, these results upgrade the coherence proxy from a single-region diagnostic to a validated applicability screen for machine-learning satellite precipitation correction.

## 4. Discussion

### *4.1. The Mechanism Purity Proposition: Reconciling Machine Learning Expectations with Physical Reality*

The TMI framework advances a counterintuitive proposition: the primary constraint on machine learning correction of satellite precipitation is not algorithmic complexity per se, but mechanism purity—the degree to which a single, coherent physical process dominates the training domain. The three empirical boundaries documented in Sections 3.3–3.5 converge on this point: correction failure typically does not occur where data are sparsest, but where the dominant mechanism is most fragmented.

The spatial boundary (Section 3.3) is the most instructive. Despite simultaneous activation of major correction levers (dem $\rho = 0.706$, tcwv $\rho = 0.875$), Central Hunan's $R^2$ degrades catastrophically (Section 3.3.2), constituting what we term mechanistic overfitting: the model migrates terrain-correction logic to stratiform-dominated grid cells, producing directionally offset and systematically positive biases. This pattern poses a direct challenge to the conventional machine learning intuition that greater feature activation invariably yields higher skill—when features encode conflicting physical signals, more activation does not necessarily translate to better performance. The Central–South Hunan $R^2$ gap (Section 3.3.2) is difficult to attribute to sampling noise and is highly consistent with mechanism fragmentation induced by mixed hill–plain terrain. For operational applications, this implies that correction skill cannot be extrapolated from elevation alone; terrain mechanism consistency must be assessed first. The subregional purity classification is defined a priori from topographic coherence and precipitation climatology (Section 2.1), independently of correction outcomes; a continuous, quantitative purity metric that is likewise independent of correction skill remains to be developed (Section 4.5).

The temporal boundary (Section 3.4) reinforces this proposition through a more subtle mechanism. The seasonal directional reversal of u10 (Section 3.4)—from positive SHAP in summer to negative values in spring—corresponds to silent failure rather than abrupt collapse: four of five variables maintain directional stability across seasons, yet u10 alone reverses, and this single-variable sensitivity introduces systematic deviation into transfer applications. The 2021–2022 reversal in RF-LR relative advantage (+0.017 in 2022, −0.017 in 2021) also suggests that correction gains are constrained by precipitation regime rather than by intrinsic model superiority alone, although the magnitude of these differences ($\pm 0.017$ $R^2$) is modest and its uncertainty is not formally quantified here. This finding cautions against the common practice of unified annual training with direct cross-seasonal deployment: when key variable polarities are season-specific, pooled training averages away actionable information and produces neutralized corrections where directional adjustment is needed.

The intensity boundary (Section 3.5) further completes the unified three-boundary interpretation: extreme precipitation failure reflects a saturation frontier rather than a classic out-of-distribution problem. The t-SNE projection (Fig. 6a) shows torrential rain samples occupying the upper tail of the intensity continuum, rather than forming fully isolated clusters; yet SHAP disorder escalates monotonically with intensity, with DEM exhibiting the largest relative amplification and IMERG the highest absolute instability (Section 3.5.2). At 50 mm $d^{-1}$ as an operational boundary for heavy precipitation, terrain–moisture synergy becomes most unstable—not because extreme precipitation is geometrically "detached" from feature space, but because the learned manifold has approached its linearizable limit.

This raises an apparent paradox that the framework must answer directly: if high-purity windows favor additive models and low-purity windows defeat all learners, where is machine learning actually necessary? The answer lies in the intermediate-purity regime—coherent terrain under transitional seasons or moderate-to-heavy intensities—where relationships are approximately, but not exactly, additive. This regime-dependence of nonlinear advantage is consistent with simplicity-bias findings in the broader machine-learning literature, where gradient-trained networks preferentially learn simple, approximately linear patterns before fitting higher-order structure or noise (Arpit et al., 2017). There, ML delivers bounded but real value: the 14% MAE reduction concentrated in the moderate-rain tail (Section 3.1.2) is operationally meaningful for flood-relevant intensities even though it leaves $R^2$ unchanged. Beyond accuracy, ML plays a second, diagnostic role: the SHAP-based boundary analysis of Sections 3.3–3.5 is possible only because a flexible learner was fitted first—an additive model cannot expose the nonlinear regime structure whose breakdown defines those boundaries. Within the TMI framework, ML is therefore neither universally necessary nor useless: it is a regime-adaptive component, deployed where its measurable gains justify the interpretability cost, withheld where a priori coherence metrics flag low purity, and exploited throughout as a diagnostic instrument.

Collectively, these three boundaries support the same judgment: in this problem, ML correction is not an unconditional universal approximator (Hornik et al., 1989), but a conditionally constrained interpolator governed by mechanism purity. The TMI framework therefore shifts research emphasis from "algorithmic competition" (RF vs. LR, deep learning vs. ensembles) toward "mechanism diagnosis": determining a priori whether a given spatiotemporal window possesses sufficient mechanism purity to warrant deployment of learned correction.

The mechanism purity concept proposed in this study differs fundamentally from established ML paradigms. Physics-informed ML embeds governing equations into the loss function to enforce physical consistency, whereas mechanism purity concerns the intrinsic consistency of physical processes within the training domain itself—the former constrains model structure, the latter constrains data-generating conditions. Domain adaptation and OOD generalization assume train-test shift can be compensated by aligning distributions; the mechanism purity introduced herein reveals that when the training domain itself contains competing mechanisms, no distributional alignment can resolve intrinsic physical conflicts. Covariate shift and concept drift address feature and conditional distribution changes separately;

mechanism purity degradation involves both simultaneously and cannot be resolved through reweighting. Regime-dependent modeling in meteorology is conceptually closest to mechanism purity, yet the TMI framework advances this by providing SHAP-based quantifiable diagnostics that transform a qualitative notion into an operable boundary identification system.

### *4.2. Implications for satellite precipitation correction methodology*

Three alternative explanations for the negligible RF–LR gap can be ruled out. First, feature quality is not the bottleneck at the aggregate scale: the ablation hierarchy (Table 1) shows that the ERA5 predictors carry substantial usable signal (+0.225 $R^2$ over the IMERG-only baseline), and the tcwv–DEM dependence structure (Fig. 2b) exhibits physically coherent feature–response relationships. Second, data volume is not limiting: with 147,660 training grid-days for five predictors, both learners are far from the data-starvation regime. Third, model capacity is not limiting either: RF with 500 trees and full feature participation has ample flexibility to exceed LR wherever exploitable nonlinear structure exists. The convergence of two fundamentally different learners on the same ceiling ($R^2 \approx 0.344$) therefore indicates that the ceiling is set by the information content of the predictors and the intrinsic stochasticity of the precipitation process—not by how flexibly a model can listen.

Our ablation results (Table 1) offer an important methodological observation: under the present sample and feature conditions, RF-Full yields only a +0.001 $R^2$ improvement over LR-Full, while amplifying bias from 0.454 (LR-Full) to 1.282 mm $d^{-1}$. The approximately 14% MAE reduction reflects improved fitting in the moderate-to-heavy rain range, yet this gain comes at the cost of distributional mean shift. The near equivalence of linear and nonlinear models in global metrics suggests that terrain–moisture relationships in this region are predominantly additive in structure.

This observation carries direct implications for the selection of nonlinear architectures. Whether to adopt more complex models should be preceded by a priori linear baseline assessment; if LR-Full already approaches the RF-Full ceiling, the marginal gains from tree-based or deep models may not justify their interpretability costs and transfer uncertainties. SHAP-based mechanistic decomposition is essential at this point: without it, RF's marginal $R^2$ advantage could easily be misinterpreted as significant nonlinear capture, when it more likely corresponds to tail error redistribution.

The dominant contribution of ERA5 atmospheric variables (+0.225 $R^2$ over the IMERG-only baseline, approximately 3.6 times the DEM marginal gain) further indicates that humidity and circulation dynamics are the primary physical levers for IMERG correction in subtropical monsoon regions. The relatively limited incremental contribution of DEM (+0.062, relative to IMERG-only) does not diminish the importance of terrain; Fig. 2b demonstrates that DEM can substantially enhance correction effects under high-moisture conditions. The more parsimonious interpretation is that DEM serves primarily as a conditional moderator in this framework—its effect depends on the moisture background rather than operating as an independent dominant predictor. Consequently, future correction schemes should prioritize the quality of reanalysis atmospheric variables; 0.25° DEM already provides effective terrain information, and higher-resolution DEM without corresponding dynamic field detail may yield diminishing returns.

The bias–variance trade-off evident in Fig. 1 and Table 4 therefore requires explicit attention in operational settings. The bias amplification introduced by RF-Full (Table 1) does not reflect a calibration failure per se; rather, it arises structurally from the elevation of moderate-intensity estimates to reduce pointwise error. In hydro-meteorological applications where bias accumulation is more consequential, RF correction may not outperform linear alternatives even when $R^2$ is superior. A practical safeguard is to report RF and linear correction results in parallel, allowing users to select the product according to their task-specific loss functions, with the selection documented as part of the operational record.

### *4.3. Physical interpretability and the SHAP diagnostic paradigm*

The core contribution of the TMI framework lies in extending SHAP from a "variable ranking tool" to a "mechanism boundary diagnostic tool." Conventional feature importance metrics (e.g., permutation importance, Gini impurity) typically compress directional and interaction information into scalar rankings, making it difficult to directly reveal the physical logic underlying correction. Our directional SHAP results (Table 2) indicate that the variable ranking (imerg > tcwv > v10 > dem > u10) is itself less informative than directional patterns: for example, v10's directional consistency attenuates from −0.782 (light rain) to −0.356 (torrential rain) (Table 2), suggesting that this wind-field correction logic degrades significantly under convective-organization-dominated conditions.

The tcwv–DEM dependence plot (Fig. 2b) further supports the moisture dependence of terrain modulation: terrain-related positive contributions strengthen primarily above ~50 kg $m^{-2}$; high-elevation grid cells (>1000 m) can correspond to higher positive SHAP (some samples exceeding +20), whereas low-elevation cells (<400 m) mostly remain near zero contribution even under high moisture. This elevation-conditioned gradient supports the interpretation that DEM × tcwv interaction primarily reflects large-scale terrain modulation rather than fine-scale terrain triggering. For resolution selection, this distinction carries direct implications: 0.25° DEM (~25–28 km) can capture regional-scale macroscopic terrain contrasts (e.g., western mountains vs. northern plains), but cannot resolve ridge-scale processes; therefore, its transferability beyond the resolution and regional scale settings of this study (e.g., smaller watershed scales) requires additional validation.

Intensity-stratified SHAP dynamics (Table 2) also reveal a more nuanced interpretability challenge: u10 undergoes directional reversal across the intensity spectrum (+0.096 in light rain, −0.207 in torrential rain). This reversal is more consistent with a shift in wind-field mechanisms across intensities than with model confusion: under low-to-moderate intensity conditions, the positive contribution of u10 can be linked to moisture transport modulation; under extreme intensity conditions, the stability of its statistical relationship declines, leading to directional flip. Based on this phenomenon, SHAP directional reversal can serve as an actionable diagnostic signal for mechanism degradation, flagging prediction scenarios that require manual review or alternative modeling strategies.

### *4.4. Limitations and boundary conditions*

Several limitations constrain the generalizability of the TMI framework. First, the spatiotemporal coverage of this study is limited: a 5-year training period (2016–2020) and 2-year testing period (2021–2022) and a spatial domain spanning the South China summer monsoon region (Hunan, Guangxi, and Guangdong; Section 3.6), both insufficient to fully sample interdecadal circulation shifts and El Niño/La Niña modulation. The temporal boundary analysis (Section 3.4) demonstrates qualitative differences in spring mechanisms (u10 reversal, v10 attenuation), yet autumn and winter precipitation regimes—where cold-front dynamics may introduce additional mechanism types—remain untested. The interannual contrast between 2021 (strong Mei-yu, stratiform-dominated) and 2022 (frequent convective transitions) provides preliminary evidence for regime-dependent transferability, but longer records are needed to establish robustness. The single-province limitation of the core analysis is addressed directly in Section 3.6, where the two-arm design—direct transfer of the Hunan-trained model, testing mechanism portability, versus local retraining, testing the learnability gate—is embedded in a leave-one-region-out scheme across Hunan, Guangxi, and Guangdong, with coherence-proxy predictions pre-registered before any model evaluation. Two outcomes of that validation qualify the present conclusions. First, the geomorphic intuition that karst terrain fragmentation implies low mechanism purity is not supported at the 0.25° daily scale: Guangxi exhibits the most coherent daily precipitation fields in the study (Moran's I = 0.773–0.787) and the highest correction efficiency, so the coherence proxy, rather than terrain intuition, must be regarded as the operative screening variable. Second, the proxy quantifies the learnability of the correction signal but not the headroom available for correction: where the satellite baseline is already comparatively accurate (coastal Guangdong), efficiency is bounded even under high coherence, and the pre-registered rule over-predicted efficiency there by 4–7 percentage points. Validation beyond the South China summer monsoon regime—cold-season precipitation and other climate–terrain settings—remains future work. More broadly, the coherence proxy introduced in Section 2.3.4 functions as an applicability screen for ML-based satellite precipitation correction: because it is computable from historical observations alone, any region can be triaged a priori as likely correctable or likely resistant before any model is trained. The pre-registered cross-regional test of Section 3.6—four of five efficiency classes predicted correctly, with a mean absolute error of 2.6 percentage points—provides the first out-of-sample validation of that screen. Like any proxy, it is fallible: spatial coherence indicates, but does not uniquely identify, mechanism singularity—a single stochastic convective regime can produce fragmented fields, and distinct mechanism mixtures can yield similar coherence values—so the screen is best applied as an ordinal triage rather than a definitive classification.

Second, the intensity boundary is diagnosed at ≥50 mm $d^{-1}$, where the sample size (n = 522 grid-days, 0.88% of test data) limits statistical power for subregional decomposition. Bootstrap validation of SHAP disorder escalation (N = 1,000, all $p < 0.001$) is robust at the aggregate level, yet the spatial variability of the 50 mm $d^{-1}$ threshold cannot be determined with current samples. Future work should stratify intensity analysis by subregion to test whether spatial and intensity boundaries interact. Until such stratified analyses are available, the saturation-frontier findings should be regarded as preliminary, aggregate-level evidence.

Third, the TMI framework is currently implemented for IMERG V07 Final correction using ERA5 atmospheric variables and SRTM DEM. The principle of mechanism purity should in principle generalize to other satellite products (e.g., GSMaP, CMORPH) and reanalysis datasets (e.g., MERRA-2, JRA-55), but specific purity thresholds and variable rankings may shift with product-specific error structures—differences that are evident even across versions of the same satellite product (Gu et al., 2026; Guo et al., 2024; Li et al., 2025; Gao et al., 2024). For example, products with stronger microwave retrieval bias may exhibit different tcwv–DEM synergy patterns, while infrared-dominated products may show heightened wind-field sensitivity. Cross-product validation is essential before operationalizing TMI diagnostics.

Fourth, our RF implementation uses default scikit-learn hyperparameters (n_estimators = 500, max_features = 1.0) without systematic tuning. While hyperparameter optimization may marginally improve $R^2$, the near-equivalence with LR-Full (Table 1) suggests that additive terrain–moisture relationships are the binding constraint rather than model capacity. We avoided hyperparameter search to prevent overfitting to the test period, yet acknowledge that ensemble methods (XGBoost, LightGBM) or deep learning architectures (CNN-LSTM hybrids) might exploit residual nonlinearities in larger datasets.

Fifth, the near-equivalence of RF-Full and LR-Full (Section 4.2) reflects the additive structure of terrain–moisture relationships in Hunan summer, not a universal failure of nonlinear ML capacity. In regions with strongly nonlinear precipitation regimes—such as the Tibetan Plateau or arid northwestern China—terrain–moisture relationships may exhibit non-negligible nonlinearity, and RF advantages over LR may be more pronounced. Our conclusion should thus be read as "linear baselines approach the performance ceiling in additive systems with high mechanism purity," rather than as a general dismissal of nonlinear models.

Sixth, three statistical caveats noted earlier bound the strength of our inferences: autocorrelated grid-day samples render the reported intervals optimistic (Section 2.3.3); subregional $R^2$ contrasts are partly confounded with precipitation variance (Section 3.3.2); and the SHAP disorder index σ scales with the predicted correction magnitude (Section 3.5.2). Event- or year-block bootstrap, variance decomposition, and normalized dispersion measures (e.g., coefficients of variation of SHAP values) are needed before these diagnostics can support stronger quantitative claims. None of these caveats alters the qualitative patterns reported.

### *4.5. Future directions: From diagnosis to proactive purity assessment*

The TMI framework suggests three directions for moving from "passive failure diagnosis" toward "proactive risk management." First, mechanism purity pre-screening: before deploying ML correction, purity metrics (e.g., SHAP directional consistency, cross-variable correlation structure, precipitation regime clustering) can be assessed to flag potential failure windows. The seasonal reversal of u10 (Table 5) and DEM disorder escalation (Fig. 6b) in this study provide actionable prototypes, yet their automation into scalable scoring systems requires cross-region and cross-season validation.

Second, adaptive model selection: rather than relying on a single annual model, one may explore model-switching strategies based on real-time purity assessment (linear versus nonlinear model switching, or season-specific model switching). The 2021–2022 reversal in algorithmic advantage (Table 4) demonstrates that no single model consistently dominates across all precipitation regimes; thus, adaptive workflows triggered by purity metrics hold practical relevance.

Third, physics-informed feature engineering: TMI boundaries identify specific physical gaps in the current predictor set. The substantial DEM disorder amplification at the intensity boundary (Section 3.5) suggests that subgrid terrain factors (slope, aspect, ridge–valley index) may compensate for information loss due to 0.25° elevation averaging. Meanwhile, the u10 reversal at the temporal boundary indicates that seasonal interaction terms (e.g., u10 × season, v10 × tcwv × season) could explicitly encode regime-specific dynamics that are diluted in annual models. Compared with simply increasing model complexity, physics-informed augmentation offers a more principled path for expanding the learnable mechanism manifold.

Moreover, the concept of mechanism purity extends beyond precipitation correction to other geophysical ML tasks where training and deployment domains are prone to divergence, such as soil moisture retrieval, land surface temperature downscaling, and atmospheric constituent estimation. The central insight of TMI is that ML effectiveness is constrained not only by data volume and model capacity, but more fundamentally by physical mechanism consistency.

A Mechanism Saturation Index (MSI) that formalizes the SHAP disorder escalation for uncertainty quantification and early warning under extreme precipitation is under development and will be reported separately.

## 5. Conclusions

This study proposes the Terrain–Moisture–Intensity (TMI) framework. Centered on mechanism purity as the core diagnostic dimension, the framework reframes satellite precipitation machine learning correction from a purely algorithmic optimization problem to one constrained by physical consistency. Based on empirical analysis of IMERG V07, SRTM DEM, and ERA5 variables for summer precipitation correction in Hunan Province, four conclusions are drawn:

(i) Under the sample and feature conditions of this study, terrain–moisture relationships in Hunan summer precipitation are predominantly additive: RF adds negligible $R^2$ over the linear baseline, and its MAE improvement comes with amplified mean bias—a structural trade-off among error metrics (Section 3.1).

(ii) Correction degradation is governed primarily by mechanism purity: spatially through mechanistic overfitting in mixed terrain, temporally through seasonal polarity shifts of key wind variables, and in intensity through a saturation frontier at extreme precipitation rather than classic out-of-distribution failure (Sections 3.3–3.5). Subregional variance and sample-size differences are partially confounded with purity in the present design (Section 4.4).

(iii) SHAP analysis serves not merely for variable ranking, but for mechanism diagnosis: directional consistency identifies where predictors lose their stable physical meaning across mechanism regimes, and the SHAP disorder index quantifies the progression of mechanism saturation (Section 4.3).

(iv) The TMI framework unifies spatial, temporal, and intensity boundaries under the principle of "mechanism consistency": machine learning correction remains stable and effective primarily under conditions of high mechanism purity. This paradigm carries direct implications for adaptive model selection in operational deployment and physics-constrained feature engineering. Pre-registered validation across Hunan, Guangxi, and Guangdong upgrades the framework from a single-region proof of concept to a quantitatively tested applicability screen: a priori coherence proxies predicted cross-regional correction efficiency with a mean absolute error of 2.6 percentage points, and the transfer-versus-retraining contrast cleanly separated mechanism mismatch from learnability (Section 3.6). Extending the screen beyond the South China monsoon domain and to cold-season regimes is the necessary next step.

**Funding**

This research did not receive any specific grant from funding agencies in the public, commercial, or not-for-profit sectors.

**Acknowledgments**

The author thanks NASA/Goddard Space Flight Center, the Copernicus Climate Change Service, and the National Climate Center of China for data support.

**CRediT authorship contribution statement**

Yi Xu: Conceptualization, Methodology, Software, Validation, Formal analysis, Investigation, Data curation, Writing – original draft, Writing – review & editing, Visualization, Project administration.

## Declaration of generative AI and AI-assisted technologies in the manuscript preparation process

During the preparation of this work the author used Kimi in order to translate and polish the language. After using this tool, the author reviewed and edited the content as needed and takes full responsibility for the content of the published article.

## Declaration of competing interest

The author declares that he has no known competing financial interests or personal relationships that could have appeared to influence the work reported in this paper.

## Data and Code Availability Statement

GPM IMERG Final Run daily precipitation data (Version 07) are available from NASA GES DISC (https://doi.org/10.5067/GPM/IMERGDF/DAY/07, Huffman et al., 2023). ERA5 hourly reanalysis data on single levels are available from the Copernicus Climate Data Store (https://doi.org/10.24381/cds.adbb2d47, Copernicus Climate Change Service, 2023). The CN05.1 gridded daily precipitation dataset (Wu & Gao, 2013; Xu et al., 2009; Wu et al., 2017) can be requested from the Nansen-Zhu International Research Centre, Institute of Atmospheric Physics, Chinese Academy of Sciences. SRTM 30 m DEM elevation data for Hunan Province are available from the Big Earth Data Science Data Center, Chinese Academy of Sciences (CASEarth) (https://doi.org/10.12237/casearth.67ad563083917d6a7fa53543).

The complete codebase for this project—including the training and ablation pipeline, the shared coherence-proxy implementation (core/morans_proxies.py), and the cross-regional validation script that generates Tables S5–S8, Figures 7–8, and Figure S2—is publicly hosted on GitHub (https://github.com/YiXuProf/tmi-framework) and archived on Zenodo (https://doi.org/10.5281/zenodo.21757101).

**Supplementary Materials**

The following supplementary materials are available online:

Table S1. K-means++ clustering evaluation (K = 2–8; Arthur and Vassilvitskii, 2007) based on standardized features (latitude, longitude, DEM, IMERG, observed precipitation). Optimal partition is determined by composite ranking of the silhouette coefficient, Calinski–Harabasz index, and Davies–Bouldin index.

Table S2. Geographic and hydroclimatic centroids of K-means clusters (K = 4). Cluster labels (Northwest/Northeast/Southwest/Central) are assigned by matching centroids to geographic templates. Observed and IMERG means confirm the Central cluster as a distinct high-precipitation island.

Fig. S1. t-SNE perplexity sensitivity analysis. Five perplexity values (5, 10, 30, 50, 100) are tested on a stratified subsample (N = 822: 522 torrential-rain events ≥ 50 mm $d^{-1}$, 300 non-torrential events < 50 mm $d^{-1}$). Stability metrics are reported in Table S3.

Table S3. t-SNE perplexity sensitivity metrics. Silhouette coefficient, centroid distance ratio (between-class versus within-class), and k=10 nearest-neighbor purity. Mean coefficient of variation across the three metrics = 0.150.

Table S4. Bootstrap verification of SHAP disorder escalation across precipitation-intensity quintiles. σ: standard deviation of SHAP values within the lowest (0–20%) and highest (80–100%) quintiles; $\Delta\sigma = \sigma_{high} - \sigma_{low}$; CI: percentile bootstrap 95% confidence interval (N = 1,000).

Table S5. Subregion coherence proxies and correction efficiency for Hunan (JJA 2021–2022): mean daily Moran's I of observed precipitation (± SE over days), observational variance, and RF correction efficiency, computed with the unified implementation released in the project repository.

Table S6. Leave-one-region-out transfer matrix: RF and LR correction efficiency (%) for all source → target province pairs (JJA 2021–2022), with day-block bootstrap 95% confidence intervals for RF.

Table S7. Pre-registered prediction check: coherence–efficiency rule fitted on the Hunan subregions (efficiency = 84.30 × Moran's I − 28.98); predicted versus realized retrained RF correction efficiency for five Guangxi and Guangdong subregions.

Table S8. Pre-registration record for the cross-regional test: rule coefficients, registration timestamp (UTC), and the class definitions used in Table S7.

Fig. S2. Maps of mean daily Moran's I of observed precipitation (JJA 2021–2022) for Hunan, Guangxi, and Guangdong.

# Supplementary Materials for

## Machine learning correction of satellite precipitation is governed by mechanism purity, not algorithmic complexity: a proof-of-concept study in Hunan, China, with pre-registered cross-regional validation

Yi Xu*

School of Information Engineering, Hunan Industry Polytechnic, Changsha, Hunan, China

*Corresponding author: Yi.Xu.Prof@outlook.com

## Table S1

*K-means clustering evaluation for K = 2–8 on standardized features (latitude, longitude, DEM, IMERG, observed precipitation). Silhouette: silhouette coefficient; Calinski-Harabasz: cluster separation index; Davies-Bouldin: cluster compactness-separation ratio; rank_sum: sum of ranks across three metrics (lower = better). The silhouette coefficient alone favors K = 2 (0.447), indicating a dominant large-scale split; the composite rank across the three indices favors K = 4 (rank_sum = 8), which we adopt for its agreement with the physically motivated four-regime structure (Section 2.1).*

| K | Silhouette | Calinski-Harabasz | Davies-Bouldin | rank_sum | Optimal |
|---|---|---|---|---|---|
| 2 | 0.447 | 15501 | 1.344 | 15 | |
| 3 | 0.282 | 19013 | 1.326 | 11 | |
| 4 | 0.278 | 19682 | 1.145 | 8 | ✓ |
| 5 | 0.278 | 18312 | 1.088 | 10 | |
| 6 | 0.286 | 17607 | 1.202 | 9 | |
| 7 | 0.268 | 16904 | 1.307 | 17 | |
| 8 | 0.282 | 16200 | 1.206 | 14 | |

## Table S2

*Geographic and hydro-climatic centroids of the K = 4 clusters. Cluster names (NW/NE/SW/Central) assigned by centroid-to-geographic-template matching. The three geographic clusters (NW/NE/SW) show similar mean precipitation (2.8–3.0 mm d$^{-1}$), whereas the Central cluster concentrates high-precipitation grid-days (obs_mean = 29.2 mm d$^{-1}$, imerg_mean = 36.3 mm d$^{-1}$) whose geographic centroid falls within the Central Hunan region. The NW, NE, and SW clusters correspond approximately to the West, North, and South Hunan subregions of the template partition, respectively, although their boundaries are data-driven and do not coincide exactly with the subjective partition; the K = 4 solution is thus supportive of, but not identical to, the four-regime structure. Because precipitation variables enter the feature set, this cluster partly reflects precipitation intensity rather than geography alone; it is therefore used as a consistency check for the subregional partition, not as independent validation (Section 3.3.2).*

| cluster_id | Zone | N | lat_mean | lon_mean | DEM_mean (m) | obs_mean (mm d⁻¹) | imerg_mean (mm d⁻¹) |
|---|---|---|---|---|---|---|---|
| 0 | NW | 20,037 | 27.960 | 110.218 | 568.1 | 2.913 | 2.759 |
| 1 | NE | 19,643 | 28.421 | 112.448 | 122.8 | 2.817 | 3.058 |
| 2 | Central | 4,388 | 27.705 | 111.474 | 370.7 | 29.210 | 36.327 |
| 3 | SW | 14,996 | 26.001 | 112.364 | 426.3 | 2.982 | 2.675 |

## Figure S1

*t-SNE perplexity sensitivity analysis. Five perplexity values (5, 10, 30, 50, 100) were tested on a stratified subsample (N = 822: 522 torrential-rain events ≥ 50 mm d⁻¹, 300 non-torrential events < 50 mm d⁻¹). Red: torrential-rain samples; blue: non-torrential samples. Stability metrics summarized in Table S3: mean coefficient of variation across silhouette coefficient, centroid separation, and nearest-neighbor purity = 0.150, confirming that the partial-overlap pattern in Fig. 6a is robust to perplexity choice.*

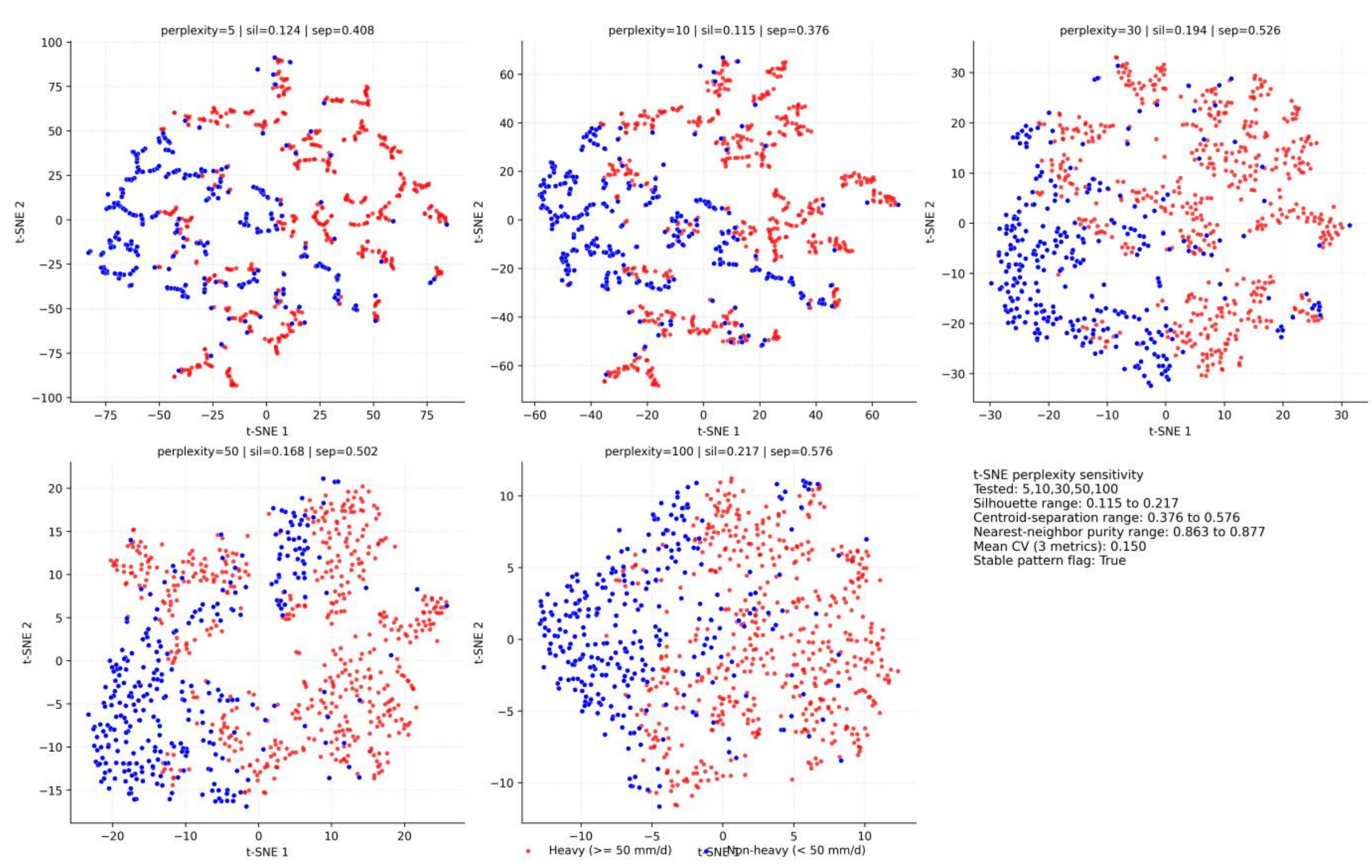

## Table S3

*t-SNE perplexity sensitivity metrics. Silhouette: silhouette coefficient for torrential vs. non-torrential binary clustering; centroid_sep: ratio of between-cluster to within-cluster centroid distance; neighbor_purity: fraction of k=10 nearest neighbors sharing the same class label. Mean CV across three metrics = 0.150.*

| perplexity | N_total | N_heavy | N_non_heavy | silhouette | centroid_sep | neighbor_purity |
|---|---|---|---|---|---|---|

| perplexity | N_total | N_heavy | N_non_heavy | silhouette | centroid_sep | neighbor_purity |
| --- | --- | --- | --- | --- | --- | --- |
| 5 | 822 | 522 | 300 | 0.124 | 0.408 | 0.863 |
| 10 | 822 | 522 | 300 | 0.115 | 0.376 | 0.869 |
| 30 | 822 | 522 | 300 | 0.194 | 0.526 | 0.875 |
| 50 | 822 | 522 | 300 | 0.168 | 0.502 | 0.877 |
| 100 | 822 | 522 | 300 | 0.217 | 0.576 | 0.869 |

## Table S4

*Bootstrap verification of SHAP disorder escalation across precipitation intensity quintiles. σ: standard deviation of SHAP values within the lowest (0–20%) and highest (80–100%) intensity quintiles. $\Delta\sigma = \sigma_{high} - \sigma_{low}$. CI: percentile bootstrap 95% confidence interval (N = 1,000). All four variables show significantly elevated disorder in the high-intensity quintile, quantifying the progressive failure boundary.*

| Feature | σ_low (0–20%) | σ_high (80–100%) | Δσ | 95% CI |
| --- | --- | --- | --- | --- |
| dem | 0.871 | 3.593 | 2.722 | [2.656, 2.794] |
| imerg | 2.604 | 7.679 | 5.075 | [4.899, 5.242] |
| tcwv | 1.813 | 6.225 | 4.412 | [4.300, 4.519] |
| v10 | 1.307 | 3.200 | 1.893 | [1.807, 1.970] |

## Table S5

A priori regime coherence and variance-free correction efficiency by subregion (test period 2021–2022). Moran's I: mean daily spatial coherence of observed precipitation (queen contiguity on the 0.25° grid), with standard error over 184 days. RMSE reduction: $(1 - RMSE_RF / RMSE_IMERG) \times 100\%$, independent of variance normalization. Central Hunan exhibits both the lowest spatial coherence and the lowest correction efficiency. Values are computed with the unified cross-regional implementation released in the project repository; the subregion ranking and the Central-Hunan deficit are unchanged from the previous version of this table.

| Subregion | Moran's I | SE | RMSE IMERG (mm d⁻¹) | RMSE RF (mm d⁻¹) | RMSE reduction (%) | Obs. variance (mm² d⁻²) |
| --- | --- | --- | --- | --- | --- | --- |
| South Hunan | 0.620 | 0.014 | 9.488 | 6.967 | 26.6 | 88.7 |
| West Hunan | 0.689 | 0.013 | 12.879 | 9.003 | 30.1 | 135.3 |
| North | 0.686 | 0.012 | 11.332 | 8.369 | 26.2 | 93.9 |
| Hunan | | | | | | |
| Central Hunan | 0.555 | 0.014 | 8.644 | 7.243 | 16.2 | 60.5 |

## Table S6

*Leave-one-region-out transfer matrix: correction efficiency (%, JJA 2021–2022) for all source → target province pairs. RF entries carry day-block bootstrap 95% confidence intervals (1,000 resamples); LR entries are point values. Diagonal (within-region) entries are optimal in every column.*

| Source → Target | RF efficiency (%) | 95% CI | LR efficiency (%) |
|---|---|---|---|
| Hunan → Hunan | 26.5 | [19.0, 34.5] | 26.5 |
| Hunan → Guangxi | 28.2 | [23.1, 32.3] | 33.6 |
| Hunan → Guangdong | 16.9 | [8.4, 25.1] | 21.4 |
| Guangxi → Hunan | 23.8 | [15.5, 32.2] | 21.5 |
| Guangxi → Guangxi | 36.5 | [29.2, 42.5] | 33.1 |
| Guangxi → Guangdong | 19.0 | [8.9, 27.1] | 20.2 |
| Guangdong → Hunan | 20.8 | [12.6, 28.7] | 19.9 |
| Guangdong → Guangxi | 26.8 | [20.9, 31.9] | 29.5 |
| Guangdong → Guangdong | 25.0 | [17.6, 31.5] | 25.2 |

## Table S7

*Pre-registered prediction check. The coherence–efficiency rule was fitted on the four Hunan subregions only (efficiency = 84.30 × Moran's I − 28.98) and locked before any Guangxi or Guangdong model was evaluated (Table S8). Classes: low < 20%, medium 20–30%, high > 30%. Result: 4/5 class hits; mean absolute error 2.6 percentage points.*

| Subregion | Moran's I | Predicted eff. (%) | Realized eff. (%) | Abs. error (pp) | Predicted class | Realized class | Hit |
|---|---|---|---|---|---|---|---|
| Guangxi karst NW | 0.787 | 37.4 | 37.9 | 0.5 | high | high | yes |
| Guangxi SE hills | 0.773 | 36.2 | 35.6 | 0.6 | high | high | yes |
| Guangdong coastal | 0.664 | 27.0 | 22.9 | 4.1 | medium | medium | yes |

| Guangdong inland hills | 0.707 | 30.6 | 23.2 | 7.4 | high | medium | no |
|---|---|---|---|---|---|---|---|
| Guangdong northern mtn. | 0.683 | 28.6 | 29.1 | 0.6 | medium | medium | yes |

## Table S8

*Pre-registration record for the cross-regional test: fitted rule coefficients, UTC registration timestamp, the record file written before any Guangxi or Guangdong model evaluation (enforced by code order), efficiency-class definitions, proxy definition, and evaluation window.*

| Item | Value |
|---|---|
| Rule | efficiency = 84.30 × Moran's I − 28.98 (fitted on Hunan subregions only) |
| Registered at (UTC) | 2026-08-07T03:49:16 |
| Record file | predictions_registered.csv — written before any Guangxi/Guangdong model evaluation, enforced by code order in xreg_validation.py |
| Class definitions | low < 20%; medium 20–30%; high > 30% |
| Proxy definition | mean daily Moran's I of CN05.1 precipitation; queen contiguity, row-standardized weights; JJA 2021–2022; persistent-finite cells |
| Evaluation window | JJA 2021–2022 (183 days for Guangxi/Guangdong; 184 days for Hunan) |

## Figure S2

*Maps of mean daily Moran's I of observed precipitation (JJA 2021–2022) for Hunan, Guangxi, and Guangdong, with subregion boundaries. Subregion-mean values are reported in Tables S5 (Hunan) and S7 (Guangxi and Guangdong).*

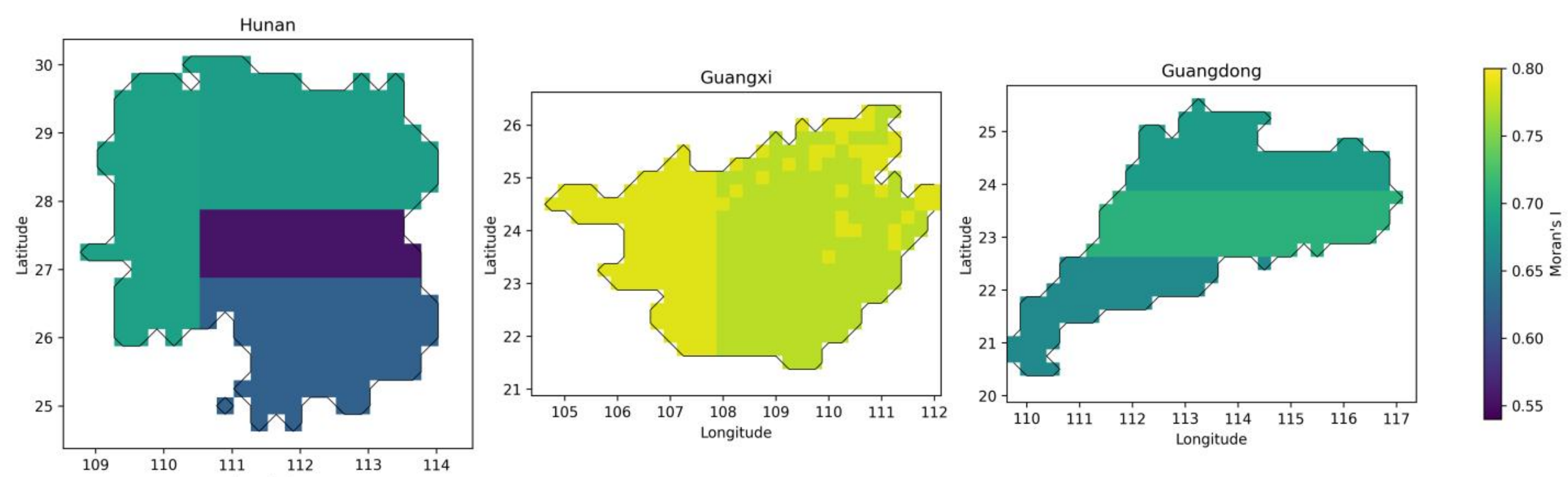